\documentclass[aps,prd,twocolumn]{revtex4-2}
\usepackage{hyperref}
\usepackage{amsmath}
\usepackage{amssymb}
\usepackage{graphicx}

\begin{document}

\title{SU(2) lattice gauge theory on a quantum annealer}

\author{Sarmed \underline{A Rahman}}
\author{Randy \underline{Lewis}}
\author{Emanuele \underline{Mendicelli}}
\author{Sarah \underline{Powell}}
\affiliation{Department of Physics and Astronomy, York University, Toronto, Ontario, M3J 1P3, Canada}

\date{August 3, 2021}

\begin{abstract}
Lattice gauge theory is an essential tool for strongly interacting non-Abelian fields, such as those in quantum chromodynamics where lattice
results have been of central importance for several decades.
Recent studies suggest that quantum computers could extend the reach of lattice gauge theory in dramatic ways,
but the usefulness of quantum annealing hardware for lattice gauge theory has not yet been explored.
In this work, we implement SU(2) pure gauge theory on a quantum annealer for lattices comprising a few plaquettes in a row with
a periodic boundary condition.
These plaquettes are in two spatial dimensions and calculations use the Hamiltonian formulation
where time is not discretized.
Numerical results are obtained from calculations on D-Wave Advantage hardware for eigenvalues, eigenvectors,
vacuum expectation values, and time evolution.
The success of this initial exploration indicates that the quantum annealer might become a useful hardware
platform for some aspects of lattice gauge theories.
\end{abstract}

\maketitle

\section{Introduction}

Lattice gauge theory is a mainstay for studies of quantum chromodynamics (QCD) and other strongly coupled gauge theories \cite{Brambilla:2014jmp}.
Significant computational resources are required, but lattice calculations
provide accurate and precise information about many of the interesting properties of nucleons and other hadrons
directly from first principles \cite{Kronfeld:2012uk}.
The lattice QCD research community has a history of evaluating each type of newly available computing hardware for its possible use \cite{Rago:2017pyb}, with
an emerging example being qubit-based computing which has produced a lot of enthusiasm and research activity \cite{Banuls:2019bmf}.
The first simulations of a lattice gauge theory on digital qubit hardware were reported in Ref.~\cite{Martinez:2016yna} for
U(1) with fermions, followed by pure gauge SU(2) \cite{Klco:2019evd}, pure gauge SU(3) \cite{Ciavarella:2021nmj}, and
SU(2) with fermions \cite{Atas:2021ext}.
Simulations of U(1) gauge theory have also been performed on analog quantum hardware \cite{Kokail:2018eiw,Mil1128,Yang:2020yer}.
The pure gauge SU(2) calculations are of particular relevance to the present work, along with many theoretical studies of Hamiltonian
SU(2) gauge theory that relate to possible qubit approaches \cite{Chandrasekharan:1996ih,Mathur:2004kr,Byrnes:2005qx,Banerjee:2012xg,Tagliacozzo:2012df,Zohar:2012xf,Zohar:2013zla,Stannigel:2013zka,Anishetty:2014tta,Zohar:2014qma,Mezzacapo:2015bra,Silvi:2016cas,Banuls:2017ena,Banerjee:2017tjn,Raychowdhury:2018tfj,Sala:2018dui,Raychowdhury:2018osk,Zohar:2019ygc,Raychowdhury:2019iki,Ji:2020kjk,Kasper:2020akk,Davoudi:2020yln,Dasgupta:2020itb,Kasper:2020owz}.
No lattice gauge theory calculations on a quantum annealer have been reported until the present work, though another group has used a quantum annealer for analyzing lattice QCD results obtained from classical computers \cite{Nguyen:2019gpo}.

A quantum annealer is a special-purpose type of qubit-based computing device \cite{PhysRevE.58.5355,FarhiGGS,FarhiGGLLP}.
Review articles can be found in Refs.~\cite{Das_2008,Albash_2018,Venegas_Andraca_2018,Hauke_2020}.
D-Wave Systems Inc.\ \cite{Dwave} has been building quantum annealers for several years, with each generation
having a larger number of qubits and increased functionality.
The current model has a quantum processing unit (QPU) with 5760 qubits, and
each qubit is connected to 15 other qubits to form what is called a Pegasus architecture.
Groups of physical qubits can function together as a single ``logical qubit,'' and these logical qubits
can communicate with all-to-all connectivity.

Instead of providing the user with a universal set of quantum gates, the quantum annealer is designed for a specific
calculation: finding the ground state of an Ising Hamiltonian (expressed here in the quadratic unconstrained binary optimization
or ``QUBO'' form)
\begin{equation}\label{eq:ising}
H(q) = \sum_{i=1}^Nh_iq_i + \sum_{i=1}^N\sum_{j=i+1}^NJ_{ij}q_iq_j
\end{equation}
where each binary $q_i$ is 0 or 1 and the user can choose any real-valued coefficients $h_i$ and $J_{ij}$.
The hardware performs its annealing by initializing the system into the ground state of a simple Hamiltonian and then moving
quasiadiabatically to the requested Ising Hamiltonian.
The Ising model might seem rather far removed from the needs of lattice QCD and too restrictive for any hope of addressing a
broad set of observables, but a goal of our paper is to show that this
quantum annealer can indeed perform a variety of calculations for a non-Abelian lattice gauge theory.
Moreover, the ability to choose directly the coefficients in Eq.~(\ref{eq:ising}) is a convenient alternative to what could otherwise
be a long sequence of quantum gates on digital hardware.

An appeal of future fault-tolerant universal quantum computers for lattice gauge theory is the potential to
open avenues that appear roadblocked to classical computing methods, particularly calculating the evolution of physics in real time
and calculating physics in an environment with nonzero baryon density \cite{Preskill:2018fag}.
With classical computers, standard lattice gauge theory algorithms rely on Markov chain Monte Carlo in Euclidean spacetime.
The use of Euclidean spacetime offers no access to real-time dynamics.
The use of Euclidean Monte Carlo means that nonzero baryon density leads to a complex-valued Monte Carlo probability distribution and
therefore a sign problem \cite{Gattringer:2016kco}.
Both of these roadblocks can be removed by using a Hamiltonian formulation, but then hardware requirements grow exponentially with the size of the Hilbert space.
This is where the hope for a quantum computer comes in: storage of the state vectors in a qubit register can scale polynomially
rather than exponentially with the system size, so combining this with an efficient quantum algorithm could lead to practical lattice gauge theory that has access
to both real-time dynamics and nonzero baryon density.

Our study does not use the qubit register in a way that achieves polynomial scaling, but the
flexibility to easily choose inputs to Eq.~(\ref{eq:ising}) does allow for significant classical preprocessing.
Therefore the system size for each quantum calculation is only a portion of the physical Hilbert space, with no involvement from any unphysical Hilbert space.
This is an approach that is well suited for present-day quantum annealers, leaving the goal of eliminating exponential scaling to be
pursued with future hardware.
Recall that a quantum annealer can be viewed as a step toward an adiabatic implementation
of universal quantum computing \cite{PhysRevA.78.012352,Vinci,PhysRevApplied.13.034037} which is equivalent to the gate-based
implementation \cite{1366223,Aharonov}, though a universal adiabatic implementation might still require more qubits than a gate-based method to achieve the equivalency.
References \cite{Chakraborty:2020uhf,Honda:2021aum} discuss adiabatic quantum computing for lattice gauge theory in a gate-based context.

Besides their connection to future universal quantum computers, quantum annealers should also be compared with classical computers.
Classical computation will remain crucial for the development of Hamiltonian lattice gauge theory methods for years to come, and quantum
annealing may be a valuable competitor for some tasks.
A scaling advantage for quantum annealing relative to path integral Monte Carlo was demonstrated on D-Wave hardware in Ref.~\cite{King}.
The scaling advantage for a D-Wave QPU relative to classical simulated annealing was demonstrated in Ref.~\cite{PhysRevX.8.031016}.
For examples of speedups attainable by quantum annealing within oracular settings, see Refs.~\cite{PhysRevA.65.042308,PhysRevLett.109.050501}.
Error mitigation is also important for maximizing the performance of quantum annealers and this is an active research area \cite{PearsonMishraHenLidar}.
In the present work we will show that, without any special accommodations for optimization or error mitigation, precise results
can be obtained for several observables in SU(2) gauge theory on small lattices.
Future studies could build on these results to learn how well quantum annealers might eventually perform relative to classical methods.

We have chosen to study the SU(2) case because it is the smallest and simplest non-Abelian gauge theory, but it is worth
recalling the long-term motivations as well.
SU(2) is a natural first step toward SU(3), which is the gauge group for QCD.
SU(2) contributes to the understanding of SU($N$) gauge theories more generally, which helps to frame QCD in a broader perspective.
SU(2) is a viable candidate for dark matter if a fermion is added \cite{Kribs:2016cew,Francis:2018xjd}.

Various aspects of SU(2) gauge theory have been studied by other researchers who are also using a Hamiltonian
formulation that can connect to a qubit implementation \cite{Chandrasekharan:1996ih,Mathur:2004kr,Byrnes:2005qx,Banerjee:2012xg,Tagliacozzo:2012df,Zohar:2012xf,Zohar:2013zla,Stannigel:2013zka,Anishetty:2014tta,Zohar:2014qma,Mezzacapo:2015bra,Silvi:2016cas,Banuls:2017ena,Banerjee:2017tjn,Raychowdhury:2018tfj,Sala:2018dui,Raychowdhury:2018osk,Zohar:2019ygc,Klco:2019evd,Raychowdhury:2019iki,Ji:2020kjk,Kasper:2020akk,Davoudi:2020yln,Dasgupta:2020itb,Kasper:2020owz,Atas:2021ext}.
Of particular interest for our work is Ref.~\cite{Klco:2019evd}, which reports the first calculation in SU(2) pure gauge theory
on a quantum computer. Specifically, IBM Q Experience gate-based hardware \cite{IBMQ} was used to compute several steps in
the time evolution of an expectation value on a two-plaquette lattice.
That work represents an important milestone for the community, and provides a context in which the results of our present study
can be assessed.

Section~\ref{sec:makeH} of the present work describes the formulation to be used, including the chosen truncation of gauge fields and the
number of plaquettes in our lattices.
The Hamiltonian matrices are constructed for these lattice systems.
Section \ref{sec:symmetries} shows how spatial symmetries can be used to block diagonalize the Hamiltonian matrices,
arriving at a form that will be used as input for the quantum annealer.
Section \ref{sec:eigenvalues} presents our use of the D-Wave quantum annealer for computing eigenvalues and eigenvectors
as a function of the gauge coupling.
Our numerical results are shown to agree with calculations from standard algorithms running on a classical computer.
In Sec.~\ref{sec:vevs} we use the D-Wave quantum annealer to calculate some vacuum expectation values as functions of the gauge
coupling, compare them with classical calculations, and use them to determine the systematic effects due to gauge truncation
and finite lattice size.
Section~\ref{sec:time} presents a method for computing time evolution and demonstrates its performance on the D-Wave hardware.
Section~\ref{sec:summary} contains a summary and outlook.
Appendix \ref{sec:derivemag} contains extra information about deriving the Hamiltonian matrix,
Appendix \ref{sec:vacblocks} displays some of the most important blocks from our block diagonalized Hamiltonian matrices,
and Appendix \ref{sec:AQAE} describes the adaptive quantum annealer eigensolver algorithm that we have developed for this project.

\section{Preparing the Hamiltonian\label{sec:makeH}}

The Hamiltonian for SU(2) lattice gauge theory was originally derived in Ref.~\cite{Kogut:1974ag}.
Follow-up discussions in the context of quantum computing can be found in Refs.~\cite{Byrnes:2005qx,Klco:2019evd,Raychowdhury:2019iki}.
Up to an overall additive constant, the Hamiltonian is
\begin{equation}\label{eq:H}
\hat H = \frac{g^2}{2}\left(\sum_{i={\rm links}}\hat E_i^2-2x\sum_{i={\rm plaquettes}}\hat\square_i\right)
\end{equation}
where $\hat E_i^2$ is the Casimir operator representing the chromoelectric field for the $i$th lattice link.
We have suppressed color indices but $\hat E_i^2$ is the sum over the three (squared) color components
of the standard SU(2) Lie algebra, $[\hat E^a,\hat E^b]=i\epsilon^{abc}\hat E^c$ as described, for example, in
Refs.~\cite{Byrnes:2005qx,Raychowdhury:2019iki}.
The plaquette operator $\hat\square_i$ in Eq.~(\ref{eq:H}) is the trace of the product of four gauge link operators in order (clockwise or counterclockwise) around the $i$th plaquette.
The gauge coupling $g$ is the only parameter in the Hamiltonian but, following Ref.~\cite{Byrnes:2005qx},
we have defined
\begin{equation}\label{eq:defx}
x \equiv \frac{2}{g^4}
\end{equation}
which will be convenient in our work.
This coefficient agrees with Ref.~\cite{Raychowdhury:2019iki} but differs from Ref.~\cite{Klco:2019evd}.
We will report energies in units of $g^2/2$ so graphs of energy versus $x$ remain bounded in the strong coupling limit, $x\to0$.

\begin{figure}
\includegraphics[width=71mm]{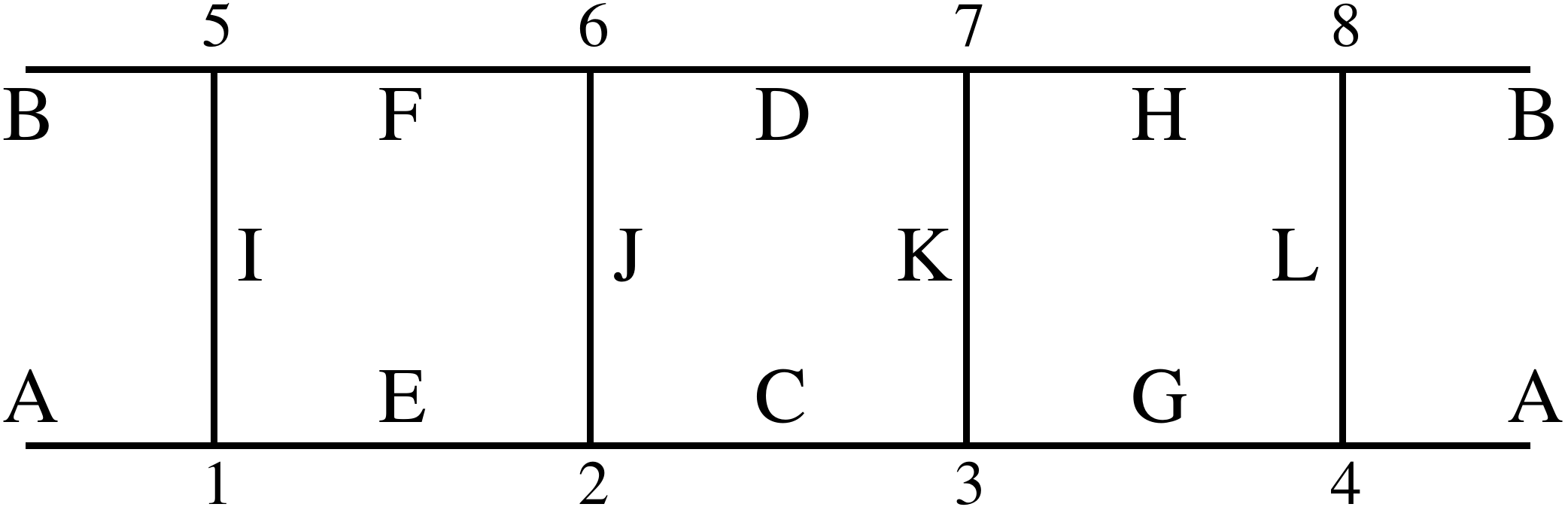}
\caption{A lattice with 12 gauge links and 4 plaquettes. It is periodic in the long direction.
         Links and sites are labeled to correspond with the calculation in Appendix \ref{sec:derivemag}.
         Each plaquette will be referenced by the number at its lower left site.\label{fig:lattice}}
\end{figure}

The Hamiltonian formalism uses a spatial lattice rather than the spacetime lattices used for standard Euclidean lattice gauge theory.
In the present work we employ a one-dimensional row of 2, 4, or 6 plaquettes with a periodic boundary in the long direction.
Figure~\ref{fig:lattice} shows the four-plaquette case.

Because SU(2) is a continuous symmetry, the Hilbert space must be truncated to encode the gauge fields into a finite qubit register.
Several formulations have been considered, and Ref.~\cite{Davoudi:2020yln} provides a useful comparison of the advantages and disadvantages for some of the leading options.
Our work begins from the so-called angular momentum formulation that was also used in Ref.~\cite{Klco:2019evd} though the
implementation is different for a quantum annealer, as will become clear in this paper.

The color state of each gauge link can be represented by a linear combination of basis states $\left|j,m,m^\prime\right>$
where $j\in\{0$, $\tfrac{1}{2}$, 1, $\tfrac{3}{2}$, \ldots\} identifies the irreducible representation (irrep) of SU(2)
while $m$ and $m^\prime$ (half integers between $\pm j$ inclusive) are the SU(2) projections at each end of the link.
Basis states for the entire lattice are products of these, so a basis state for Fig.~\ref{fig:lattice} can be written as
\begin{equation}
\left|\psi\right> = \left|j_A,m_A,m_A^\prime\right>\left|j_B,m_B,m_B^\prime\right>\ldots\left|j_L,m_L,m_L^\prime\right> \,.
\end{equation}
Color conservation (and the absence of fermions) requires that the three links arriving at any lattice site must form an
SU(2) singlet, which corresponds to Gauss's law.

To apply the Hamiltonian from Eq.~(\ref{eq:H}) to any basis state we need to consider both chromoelectric and plaquette terms.
The chromoelectric contribution is easy to evaluate \cite{Byrnes:2005qx} and for Fig.~\ref{fig:lattice} we obtain
\begin{equation}\label{eq:HE}
\left<\psi\right|\sum_i\hat E_i^2\left|\psi\right> = \sum_{i=A}^Lj_i(j_i+1) \,.
\end{equation}
Notice that these terms are on the diagonal; a matrix element between unequal states would vanish.
The plaquette contribution is a bit more involved but
we provide a derivation in Appendix \ref{sec:derivemag}.
For plaquette 1 of Fig.~\ref{fig:lattice}, the result is
\begin{align}
& \left<\psi_{\rm final}\right|\hat\square_1\left|\psi_{\rm initial}\right> \nonumber \\
 =& (-1)^{j_A+J_E+j_I}\sqrt{(2j_I+1)(2J_E+1)}\left\{\begin{array}{ccc} j_A & j_E & j_I \\ \tfrac{1}{2} & J_I & J_E \end{array}\right\} \nonumber \\
 & (-1)^{j_C+J_E+j_J}\sqrt{(2j_E+1)(2J_J+1)}\left\{\begin{array}{ccc} j_C & j_E & j_J \\ \tfrac{1}{2} & J_J & J_E \end{array}\right\} \nonumber \\
 & (-1)^{j_D+J_F+j_J}\sqrt{(2j_J+1)(2J_F+1)}\left\{\begin{array}{ccc} j_D & j_F & j_J \\ \tfrac{1}{2} & J_J & J_F \end{array}\right\} \nonumber \\
 & (-1)^{j_B+J_F+j_I}\sqrt{(2j_F+1)(2J_I+1)}\left\{\begin{array}{ccc} j_B & j_F & j_I \\ \tfrac{1}{2} & J_I & J_F \end{array}\right\} \nonumber \\
\label{eq:oneplaq}
\end{align}
where $j_i$ and $J_i$ are the SU(2) irreps in $\left|\psi_{\rm initial}\right>$ and $\left|\psi_{\rm final}\right>$, respectively.
The 6j symbols are merely square roots of ratios and are provided in Appendix \ref{sec:derivemag}.
They enforce Gauss's law automatically.
Notice that $\left|\psi_{\rm initial}\right>$ and $\left|\psi_{\rm final}\right>$ will never be the same state because
applying a plaquette operator necessarily changes each of those four gauge links by $\pm\tfrac{1}{2}$.
Therefore all plaquette terms are off diagonal.

The result in Eq.~(\ref{eq:oneplaq}) applies to a lattice of any length, not just Fig.~\ref{fig:lattice}, because only gauge links
comprising or touching the active plaquette (E,F,I,J comprise and A,B,C,D touch) are involved in the calculation.
Our Eq.~(\ref{eq:oneplaq}) agrees with the expression given in Ref.~\cite{Klco:2019evd}.

Because all sums over the projections $m$ and $m^\prime$ have already been performed to arrive at Eq.~(\ref{eq:oneplaq}),
any state of the lattice depends only on the irrep values $j_i$.
It is now straightforward to calculate each entry in the Hamiltonian matrix for any one-dimensional periodic lattice of plaquettes.
Step 1: Begin with the bare vacuum (all $j$ values set to zero) and apply any number of plaquette operators to create all
possible new basis states below a chosen maximum $j$ value.
Step 2: Use Eqs.~(\ref{eq:H}), (\ref{eq:HE}), and (\ref{eq:oneplaq}) to calculate every entry in the Hamiltonian matrix.

\begin{figure}
\includegraphics[width=6cm]{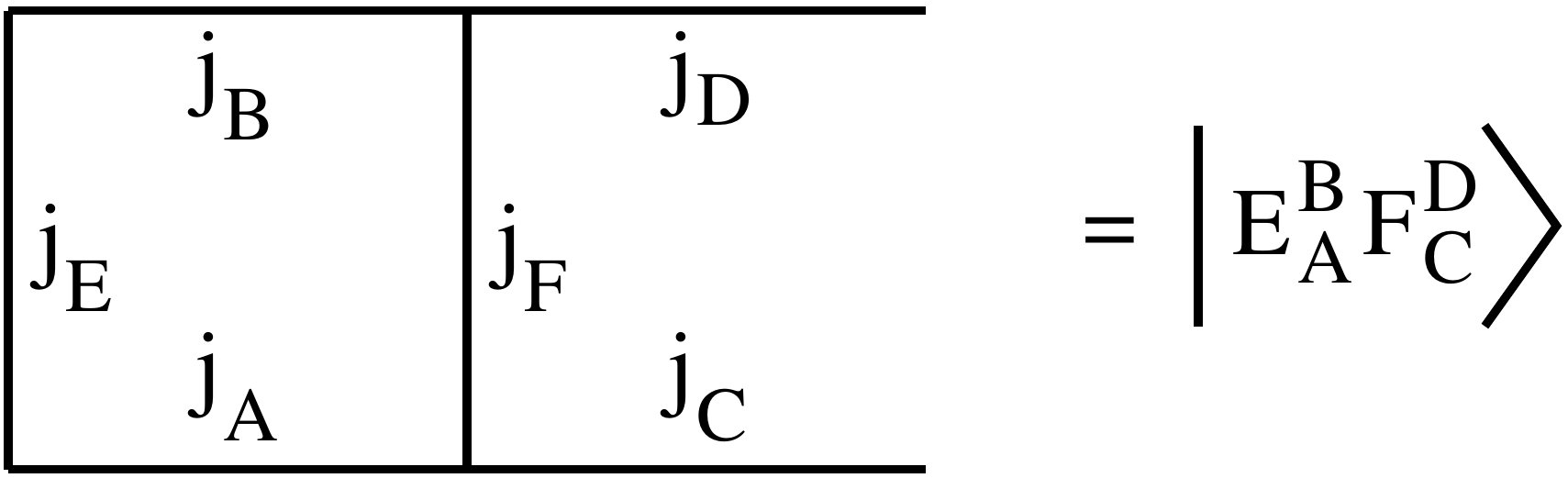}
\caption{The basis states of a lattice are represented in ket notation by using the multiplicities, $A\equiv2j_A+1$,
         $B\equiv2j_B+1$, $C\equiv2j_C+1$, etc., in a pattern that matches the layout of the spatial lattice.\label{fig:notation}}
\end{figure}

Our first example is the case considered in Ref.~\cite{Klco:2019evd},
namely the two-plaquette lattice with each gauge link truncated by $j\leq\tfrac{1}{2}$, where the Hamiltonian matrix is
\begin{equation}\label{eq:4x4}
H = 
\frac{g^2}{2}\left(
\begin{array}{cccc}
0 & -2x & -2x & 0 \\
-2x & 3 & 0 & -\frac{x}{2} \\
-2x & 0 & 3 & -\frac{x}{2} \\
0 & -\frac{x}{2} & -\frac{x}{2} & 3
\end{array}
\right)
\begin{array}{c}
\left|1_1^11_1^1\right> \\
\left|2_2^22_1^1\right> \\
\left|2_1^12_2^2\right> \\
\left|1_2^21_2^2\right>
\end{array} \,.
\end{equation}
The states listed to the right of the matrix identify the rows (and corresponding columns) by using the notation of Fig.~\ref{fig:notation}.
Our numerical studies will also go beyond this case in two ways, namely by increasing the length of the lattice and by increasing the cutoff on $j$.
Instead of displaying these larger Hamiltonian matrices explicitly, their sizes are listed in Table~\ref{tab:sizes}.
\begin{table}
\caption{The five physics systems studied in this work are identified by the number of plaquettes in the lattice, $N_{\rm plaq}$,
         and the irrep truncation, $j_{\rm max}$. For each system, this table shows the size of the original Hamiltonian and the size
         of the largest block that remains after block diagonalization.\label{tab:sizes}}
\begin{ruledtabular}
\begin{tabular}{cccc}
$N_{\rm plaq}$ & $j_{\rm max}$ & Size of $H$ & Size of vacuum block \\
\hline
2 & 1/2 &  4$\times$4  &  3$\times$3  \\
4 & 1/2 & 16$\times$16 &  6$\times$6  \\
6 & 1/2 & 64$\times$64 & 13$\times$13 \\
2 &  1  & 27$\times$27 & 14$\times$14 \\
2 & 3/2 & 95$\times$95 & 36$\times$36
\end{tabular}
\end{ruledtabular}
\end{table}
\begin{table}
\caption{The 64 basis states for the lattice with $N_{\rm plaq}=6$ and $j_{\rm max}=\tfrac{1}{2}$.
         The notation $\left|Q_i^{(j)}\right>$ denotes a state in the $i$th set where the excitation has been translated
         to the right by $j$ sites.\label{tab:6plaq}}
\begin{ruledtabular}
\begin{tabular}{ccc}
Set label & Number of states & The starting state \\
\hline
 1 & 1 & $\left|Q_1^{(0)}\right> = \left|1_1^11_1^11_1^11_1^11_1^11_1^1\right>$ \\
 2 & 6 & $\left|Q_2^{(0)}\right> = \left|2_2^22_1^11_1^11_1^11_1^11_1^1\right>$ \\
 3 & 6 & $\left|Q_3^{(0)}\right> = \left|2_2^21_2^22_1^11_1^11_1^11_1^1\right>$ \\
 4 & 6 & $\left|Q_4^{(0)}\right> = \left|2_2^22_1^12_2^22_1^11_1^11_1^1\right>$ \\
 5 & 3 & $\left|Q_5^{(0)}\right> = \left|2_2^22_1^11_1^12_2^22_1^11_1^1\right>$ \\
 6 & 6 & $\left|Q_6^{(0)}\right> = \left|2_2^21_2^21_2^22_1^11_1^11_1^1\right>$ \\
 7 & 6 & $\left|Q_7^{(0)}\right> = \left|2_2^21_2^22_1^12_2^22_1^11_1^1\right>$ \\
 8 & 6 & $\left|Q_8^{(0)}\right> = \left|2_2^22_1^12_2^21_2^22_1^11_1^1\right>$ \\
 9 & 2 & $\left|Q_9^{(0)}\right> = \left|2_2^22_1^12_2^22_1^12_2^22_1^1\right>$ \\
10 & 6 & $\left|Q_{10}^{(0)}\right> = \left|2_2^21_2^21_2^21_2^22_1^11_1^1\right>$ \\
11 & 6 & $\left|Q_{11}^{(0)}\right> = \left|2_2^21_2^21_2^22_1^12_2^22_1^1\right>$ \\
12 & 3 & $\left|Q_{12}^{(0)}\right> = \left|2_2^21_2^22_1^12_2^21_2^22_1^1\right>$ \\
13 & 6 & $\left|Q_{13}^{(0)}\right> = \left|2_2^21_2^21_2^21_2^21_2^22_1^1\right>$ \\
14 & 1 & $\left|Q_{14}^{(0)}\right> = \left|1_2^21_2^21_2^21_2^21_2^21_2^2\right>$ \\
\hline
  & Total=64
\end{tabular}
\end{ruledtabular}
\end{table}

\section{Applying spatial symmetries}\label{sec:symmetries}

Interchanging the states $\left|2_2^22_1^1\right>$ and $\left|2_1^12_2^2\right>$ in Eq.~(\ref{eq:4x4}) would leave the Hamiltonian matrix invariant.
This is a clue to an easy block diagonalization,
\begin{equation}\label{eq:3x3}
H = 
\frac{g^2}{2}\!\!\left(
\begin{array}{ccc|c}
0 & \!\!\!-2\sqrt{2}x\!\!\! & 0 & 0 \\
\!\!\!-2\sqrt{2}x\!\!\! & 3 & -\frac{x}{\sqrt{2}} & 0 \\
0 & -\frac{x}{\sqrt{2}} & 3 & 0 \\
\hline
0 & 0 & 0 & 3
\end{array}
\right)
\begin{array}{c}
\left|1_1^11_1^1\right> \\
\tfrac{1}{\sqrt{2}}\left(\left|2_2^22_1^1\right>+\left|2_1^12_2^2\right>\right) \\
\left|1_2^21_2^2\right> \\
\tfrac{1}{\sqrt{2}}\left(\left|2_2^22_1^1\right>-\left|2_1^12_2^2\right>\right)
\end{array}.
\end{equation}
For any matrix, performing block diagonalization is valuable because then each block can be submitted to the quantum
annealer separately. This reduces the qubit requirements.
It also allows the quantum annealer to provide the ground state of each block
instead of only providing the ground state of the original matrix.

To generalize the block diagonalization procedure systematically to larger Hamiltonian matrices,
we will apply three symmetries to the original lattice basis states: top-to-bottom reflection,
left-to-right reflection at a symmetry point, and spatial translation in the periodic (long) direction.
It is natural to describe spatial translation by using momentum states via $e^{ipx}$ but recall that the coefficients in Eq.~(\ref{eq:ising})
need to be real, so a modified approach will be used.

As an example of applying spatial symmetries, consider the lattice with six plaquettes and $j_{\rm max}=\tfrac{1}{2}$.
The Hamiltonian matrix is 64$\times$64 and will not be displayed explicitly here, but we do provide the complete list of 64
basis states in Table~\ref{tab:6plaq}.
These 64 states are collected into 14 sets according to spatial translation symmetry.
The notation $\left|Q_i^{(j)}\right>$ denotes a state in the $i$th set where the excitation has been translated
to the right by $j$ sites, so reading $\left|Q_2^{(0)}\right> = \left|2_2^22_1^11_1^11_1^11_1^11_1^1\right>$ from the
table immediately tells us that $\left|Q_2^{(3)}\right>=\left|1_1^11_1^11_1^12_2^22_1^11_1^1\right>$.

Notice that all 64 basis states are symmetric under a top-to-bottom reflection.
That will not generally be the case for $j_{\rm max}>\tfrac{1}{2}$, but it is true here.
If we had chosen $N_{\rm plaq}=2$ and $j_{\rm max}=1$ as our example then the states $\left|1_3^11_3^1\right>$ and
$\left|1_1^31_1^3\right>$ would have been present.
Block diagonalization would have been accomplished by replacing those two states with the linear combinations that are
positive and negative under the reflection, i.e., $\frac{1}{\sqrt{2}}\left(\left|1_3^11_3^1\right>\pm\left|1_1^31_1^3\right>\right)$.

The second symmetry to consider for the basis states in Table~\ref{tab:6plaq} is left-to-right reflection.
Notice that most basis states are symmetric under a left-to-right reflection for some reflection point, but states in
sets 7 and 8 are not.
Therefore block diagonalization requires replacing pairs of those states with their symmetric and antisymmetric combinations,
\begin{equation}
\left\{\left|Q_7^{(j)}\right>,\left|Q_8^{(j)}\right>\right\} ~~~\to~~~
\frac{1}{\sqrt{2}}\left(\left|Q_7^{(j)}\right> \pm \left|Q_8^{(j)}\right>\right) \,.
\end{equation}

The third (and final) symmetry to consider is spatial translation.
Invariance under spatial translation leads to conservation of linear momentum, so our Hamiltonian can be block diagonalized
into blocks of definite momentum.
The six allowed momenta on the six-plaquette lattice are $p=-2\pi/3$, $-\pi/3$, $0$, $\pi/3$, $2\pi/3$, and $\pi$
so within the 14 sets of Table~\ref{tab:6plaq} we should multiply each state by $e^{ipx}$ where $x$ is the integer
location (i.e.\ the plaquette number) assigned to the excitation in that state.
Only differences in $x$ really matter because any extra offset is an irrelevant overall phase.
To maintain real coefficients in Eq.~(\ref{eq:H}) we can use the real and imaginary parts separately by introducing the
simple factors from Table~\ref{tab:6mom}.
For the set in Table~\ref{tab:6plaq} that contains two basis states, the states to use for block diagonalizing are
\begin{eqnarray}
P_9^{(0)} &=& \frac{1}{\sqrt{2}}\left(Q_9^{(0)} + Q_9^{(1)}\right), \nonumber \\
P_9^{(1)} &=& \frac{1}{\sqrt{2}}\left(Q_9^{(0)} - Q_9^{(1)}\right),
\end{eqnarray}
which correspond respectively to $p=0$ and $p=\pi$.
For the sets in Table~\ref{tab:6plaq} that contain three basis states, the states to use for block diagonalizing are
\begin{eqnarray}
P_i^{(0)} &=& \frac{1}{\sqrt{3}}\left(Q_i^{(0)} + Q_i^{(1)} + Q_i^{(2)}\right), \nonumber \\
P_i^{(1)} &=& \frac{1}{\sqrt{6}}\left(2Q_i^{(0)} - Q_i^{(1)} - Q_i^{(2)}\right), \nonumber \\
P_i^{(2)} &=& \frac{1}{\sqrt{2}}\left(Q_i^{(1)} - Q_i^{(2)}\right),
\end{eqnarray}
which correspond respectively to $p=0$, $p=\pm2\pi/3$ real, and $p=\pm2\pi/3$ imaginary.
For the sets in Table~\ref{tab:6plaq} that contain six basis states, the states to use for block diagonalizing are
\begin{eqnarray}
P_i^{(0)} &=& \frac{1}{\sqrt{6}}\left(Q_i^{(0)} \!+\! Q_i^{(1)} \!+\! Q_i^{(2)} \!+\! Q_i^{(3)} \!+\! Q_i^{(4)} \!+\! Q_i^{(5)}\right), \nonumber \\
P_i^{(1)} &=& \frac{1}{2\sqrt{3}}\left(2Q_i^{(0)} \!+\! Q_i^{(1)} \!-\! Q_i^{(2)} \!-\! 2Q_i^{(3)} \!-\! Q_i^{(4)} \!+\! Q_i^{(5)}\right), \nonumber \\
P_i^{(2)} &=& \frac{1}{2}\left(Q_i^{(1)} \!+\! Q_i^{(2)} \!-\! Q_i^{(4)} \!-\! Q_i^{(5)}\right), \nonumber \\
P_i^{(3)} &=& \frac{1}{2\sqrt{3}}\left(2Q_i^{(0)} \!-\! Q_i^{(1)} \!-\! Q_i^{(2)} \!+\! 2Q_i^{(3)} \!-\! Q_i^{(4)} \!-\! Q_i^{(5)}\right), \nonumber \\
P_i^{(4)} &=& \frac{1}{2}\left(Q_i^{(1)} \!-\! Q_i^{(2)} \!+\! Q_i^{(4)} \!-\! Q_i^{(5)}\right), \nonumber \\
P_i^{(5)} &=& \frac{1}{\sqrt{6}}\left(Q_i^{(0)} \!-\! Q_i^{(1)} \!+\! Q_i^{(2)} \!-\! Q_i^{(3)} \!+\! Q_i^{(4)} \!-\! Q_i^{(5)}\right),
\end{eqnarray}
which correspond respectively to $p=0$, $p=\pm\pi/3$ real, $p=\pm\pi/3$ imaginary, $p=\pm2\pi/3$ real, $p=\pm2\pi/3$
imaginary, and $p=\pi$.
\begin{table}
\caption{The six momenta available on a six-plaquette lattice, and the corresponding sines and cosines.\label{tab:6mom}}
\begin{ruledtabular}
\begin{tabular}{c|cccccc}
$p$ & $-2\pi/3$ & $-\pi/3$ & $0$ & $\pi/3$ & $2\pi/3$ & $\pi$ \\
\hline
$\cos p$ & $-1/2$ & $1/2$ & $1$ & $1/2$ & $-1/2$ & $-1$ \\
$\sin p$ & $-\sqrt{3}/2$ & $-\sqrt{3}/2$ & $0$ & $\sqrt{3}/2$ & $\sqrt{3}/2$ & $0$
\end{tabular}
\end{ruledtabular}
\end{table}

Our example of $N_{\rm plaq}=6$ and $j_{\rm max}=\tfrac{1}{2}$ has now been block diagonalized.
The original 64$\times$64 Hamiltonian matrix has become two $p=0$ blocks (one 13$\times$13 and one 1$\times$1),
a $p=\pm\pi/3$ block that is 18$\times$18, a $p=\pm2\pi/3$ block that is 22$\times$22, 
and a pair of $p=\pi$ blocks (one 7$\times$7 and one 3$\times$3).
The 13$\times$13 block is displayed explicitly in Appendix \ref{sec:vacblocks}.

Because the $p=\pm\pi/3$ and $p=\pm2\pi/3$ blocks each contain forward and backward momenta, their spectra are filled with
pairs of degenerate eigenvalues.
We can break each of these blocks into two separate blocks by implementing a $\pm\pi/3$ rotation on pairs of states,
so the 18$\times$18 block becomes a pair of 9$\times$9 blocks, and similarly the 22$\times$22 block becomes a pair of 11$\times$11 blocks.

Classical computing can readily apply spatial symmetries according to the method discussed in this section.
Table~\ref{tab:sizes} lists the size of the largest block obtained after block diagonalization for the physics systems to be
studied in the present work.
Notice that the largest block is always the one containing the bare vacuum state $\left|1_1^11_1^1\ldots\right>$ and this block
will be of particular interest for computing vacuum expectation values, but every block can be implemented on a quantum annealer to
obtain the smallest eigenvalue and its eigenvector.

\section{Computing eigenvalues and eigenvectors\label{sec:eigenvalues}}

The variational method is well known from quantum mechanics as a way to approximate the ground-state wave function and ground-state energy for a given Hamiltonian.
By varying the parameters contained in the user's trial wave function, the best approximation having that particular form can
be found.
A more general approach is taken here.
Specifically, the complete vector space (without choosing any trial wave function) will be discretized in an unbiased way and
provided to the quantum annealer, which will then find the desired minimum.

In the standard variational method, the expectation value of a Hamiltonian $H$ for any proposed state $\left|\psi\right>$ bounds the ground-state energy $E_0$ according to
\begin{equation}\label{eq:VarMeth}
E_0 \leq \frac{\left<\psi\right|H\left|\psi\right>}{\left<\psi|\psi\right>}
\end{equation}
with the equality being approached as $\left|\psi\right>$ approaches the true ground state.
Notice that for the overly simplistic proposal that each entry in the vector representing $\left|\psi\right>$ is either 0 or 1, we arrive
at the Ising ground state of Eq.~(\ref{eq:ising}) which is solved by a quantum annealer.
[This is true because $q_i^2=q_i$ in Eq.~(\ref{eq:ising}).]
The general algorithm to be used in our work emerges by applying two extensions.
First, a robust numerical implementation needs to handle the possibility of a null vector ($q_i$=0 $\forall\,i$).
Second, practical applications need an implementation that can consider
any proposed state $\left|\psi\right>$ without restriction to binary entries of 0 and 1.
The authors of Refs.~\cite{Teplukhin_2019,TeplukhinNature,TeplukhinPCCP,teplukhin2021sampling}
have documented an explicit description of this general algorithm and named it the quantum annealer eigensolver (QAE).
They have found the QAE to be effective for chemistry calculations on a quantum annealer, and it is also successful
for lattice gauge theory as will be demonstrated presently.

The null vector is avoided by adding a penalty term as follows:
\begin{equation}\label{eq:F}
\left<\psi\right|H\left|\psi\right> \to F \equiv \left<\psi\right|H\left|\psi\right> - \lambda\left<\psi|\psi\right>
\end{equation}
where the parameter $\lambda$ in the penalty term is adjusted by the user.
For a small example, take $H$ to be the 3$\times$3 block in Eq.~(\ref{eq:3x3}) and use three binary variables $(q_1,q_2,q_3)$
to represent $\left|\psi\right>$.
\begin{table}
\caption{For each of the eight combinations of binary values, this table provides the normalization
(squared magnitude of the vector) and the value of $F$ from Eq.~(\ref{eq:F}).\label{tab:lambda}}
\begin{ruledtabular}
\begin{tabular}{ccc|cc}
$q_1$ & $q_2$ & $q_3$ & $\left<\psi|\psi\right>$ & $F$ \\
\hline
0 & 0 & 0 & 0 & 0 \\
0 & 0 & 1 & 1 & $3-\lambda$ \\
0 & 1 & 0 & 1 & $3-\lambda$ \\
0 & 1 & 1 & 2 & $6-\sqrt{2}x-2\lambda$ \\
1 & 0 & 0 & 1 & $-\lambda$ \\
1 & 0 & 1 & 2 & $3-2\lambda$ \\
1 & 1 & 0 & 2 & $3-4\sqrt{2}x-2\lambda$ \\
1 & 1 & 1 & 3 & $6-5\sqrt{2}x-3\lambda$
\end{tabular}
\end{ruledtabular}
\end{table}
All options for $F$ are listed in Table~\ref{tab:lambda} but let us begin with the simple case of $x=0$.
Without the penalty term, there are two options for getting the minimum $F$ and one of those is the unwanted null vector.
Keeping the penalty term means any choice $0<\lambda<3$ will provide the single normalizable state vector with the
correct minimum energy.
In practice there is no need to construct the explicit table because
if the null vector appears then the user can scan a few $\lambda$ values with the quantum annealer to
find the transition point where the null vector no longer appears.
The appropriate range for $\lambda$ is always adjacent to that transition point.

The extension beyond merely 0 and 1 entries in the state vector is accomplished by using multiple binary variables to construct
a fixed-point representation.
The $i$th entry in the proposed vector state is
\begin{equation}\label{eq:defpsi}
a_i = -q_K^i + \sum_{k=1}^{K-1}\frac{q_k^i}{2^{K-k}}
\end{equation}
where $K$ is the number of binary variables used for that entry.
Notice that the $a_i$ values are evenly spaced within $[-1,1)$.
On the quantum annealer, one logical qubit is used for each binary variable $q_i$ so finding the ground state for
an $N\times N$ matrix will use $NK$ logical qubits.
Increasing $K$ will increase the precision of the resulting eigenvalue and eigenvector.

To summarize, the state $\left|\psi\right>$ in Eq.~(\ref{eq:VarMeth}) is represented by a unit vector which is $(a_1,a_2,\ldots)$ from
Eq.~(\ref{eq:defpsi}) divided by its norm.
The original Hamiltonian is used without change.
See Appendix \ref{sec:AQAE} for our implementation.

Calculations on a D-Wave quantum annealer are performed by writing \textsc{python} codes that call D-Wave's \textsc{ocean} software suite \cite{Ocean}.
\textsc{ocean} provides the user with various options to explore optimizations and refinements,
including the ability to adjust the annealing schedule which defines how the hardware makes the quasiadiabatic
transition from its initial to final Hamiltonian.
The default annealing time is 20 microseconds.
The default annealing schedule during that 20 microseconds is described in
Ref.~\cite{DWaveManual}.
We have confirmed that acceptable results are obtained for the present project by using the default time and default
schedule.  The only hardware parameter we need to adjust is called the chain strength.

A chain is a set of physical qubits within the D-Wave hardware that is used to represent one logical qubit.
The length of each chain depends on which connections are required between this particular logical qubit and others.
We allow the \textsc{ocean} software to automatically perform the embedding of physical qubits into logical qubits, but we must
adjust the chain strength. If the strength is too low then the physical qubits within a logical qubit can disagree with
one another and lead to ambiguous physics output.
If the chain strength is too high then it competes with the physics terms in the intended calculation and puts a bias on the physics output.
The \textsc{ocean} software reports every chain breaking event and this has allowed us to easily tune the chain strength to be within an
acceptable range for all of our calculations.

The chain strength is implemented ``behind the scenes'' by the D-Wave system.
Every pair of physical qubits in a chain has an implicit Hamiltonian term of the form
$\delta H = -J_{\rm chain}\sigma^z_j\sigma^z_k$
where $\sigma^z$ is the Pauli matrix and a subscript identifies a specific physical qubit.
The coefficient $J_{\rm chain}$ is the chain strength.
Increasing the value of $J_{\rm chain}$ increases the probability that the two qubits will be aligned \cite{McGeoch}.
For the present study, typical values are between about 1 and 5, tuned at about O(10\%).

\begin{figure}
\includegraphics[width=86mm]{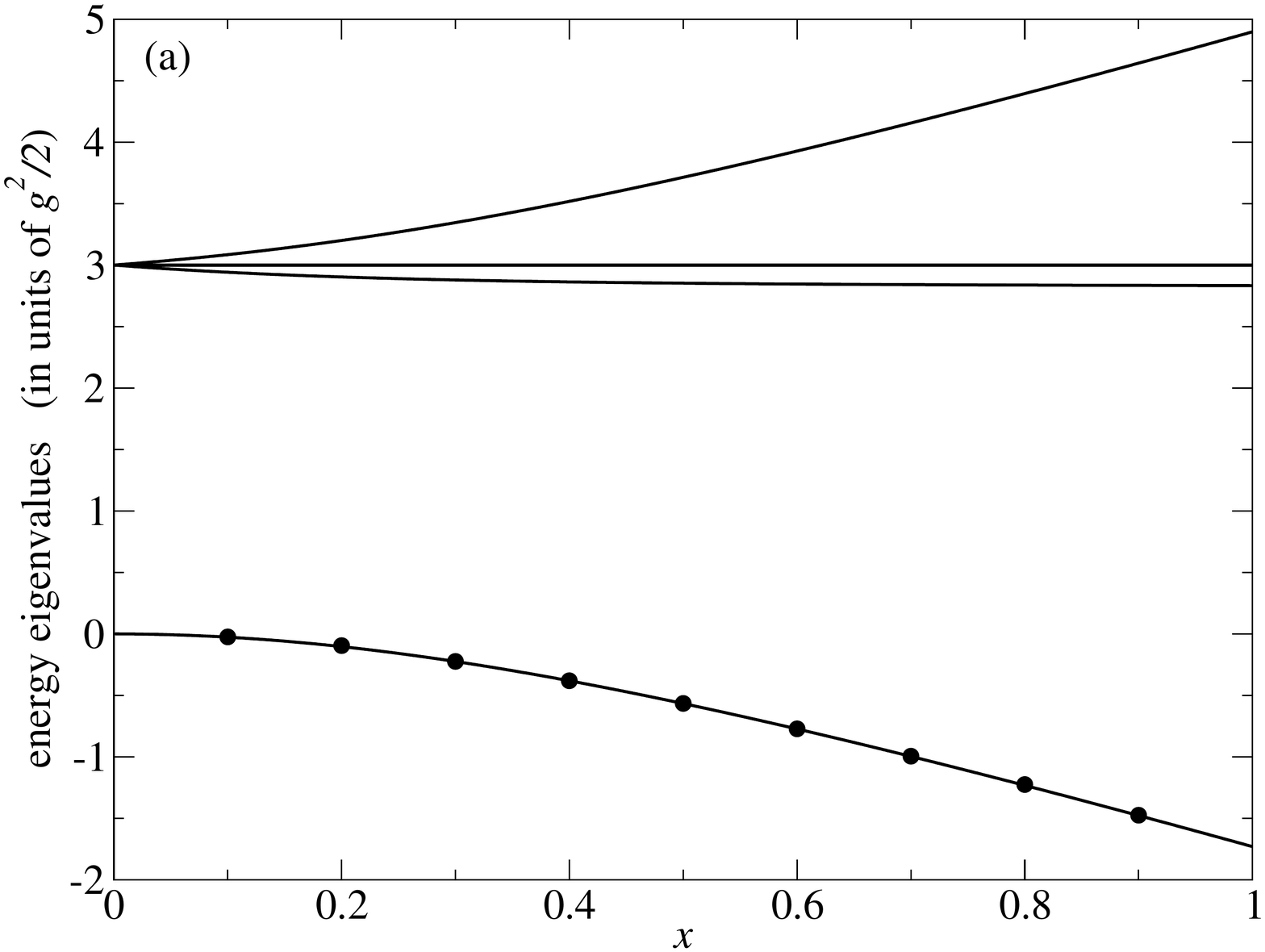}
\includegraphics[width=86mm]{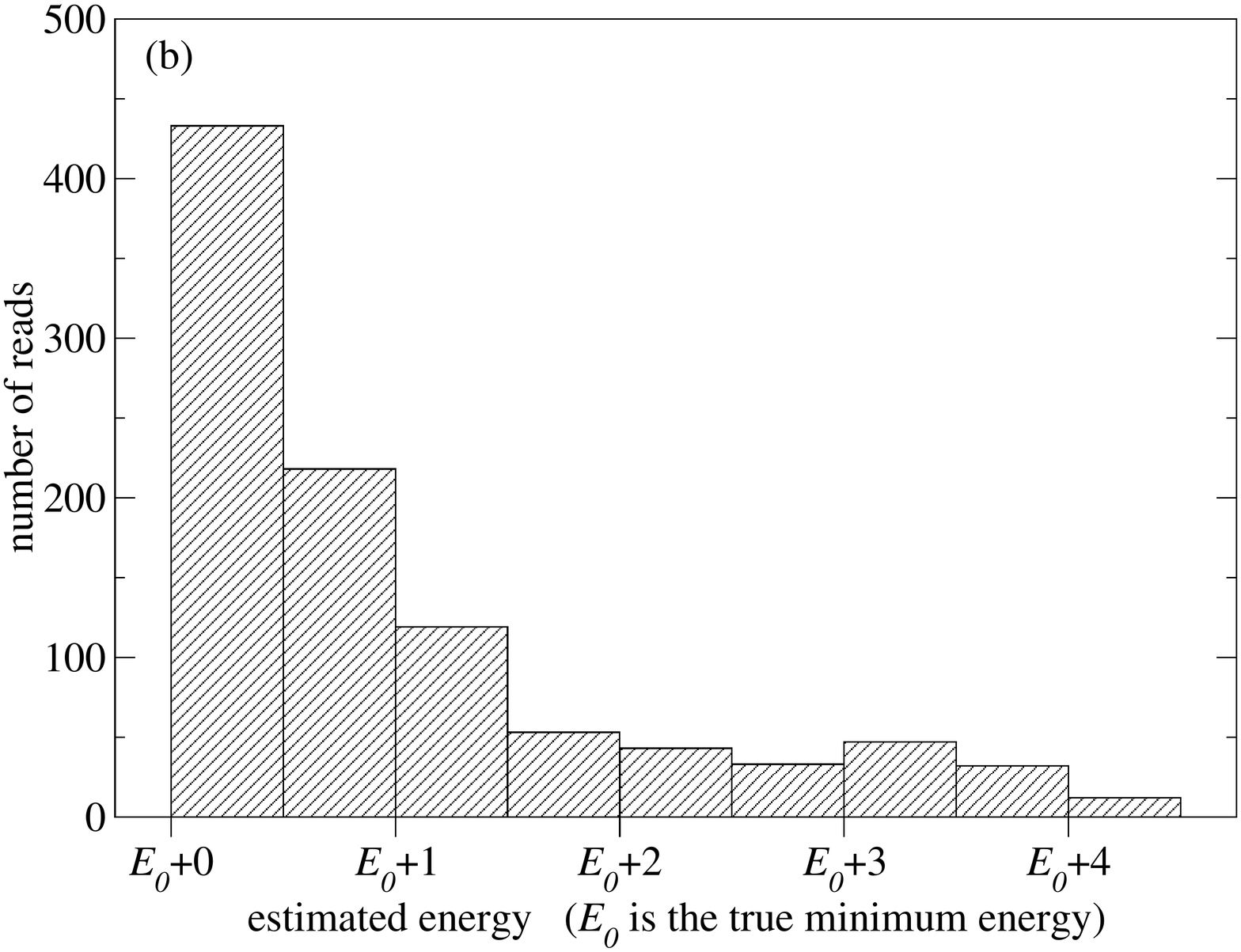}
\caption{(a) Data points show the ground-state eigenvalue for a two-plaquette lattice with $j_{\rm max}=\tfrac{1}{2}$, computed
             on the D-Wave Advantage quantum annealer as a function of the gauge coupling $x$ of Eq.~(\ref{eq:defx}).
             The four curves are the four eigenvalues of Eq.~(\ref{eq:3x3}) computed classically by exact diagonalization.
         (b) The distribution of energies obtained for 990 reads at $x=0.5$.\label{fig:3x3}}
\end{figure}
Figure~\ref{fig:3x3}(a) shows that the ground-state eigenvalue of Eq.~(\ref{eq:3x3}) calculated on the D-Wave Advantage quantum annealer is in agreement with classical exact diagonalization of Eq.~(\ref{eq:3x3}).
To make this graph, the coefficient of each basis state for the 3$\times$3 block of Eq.~(\ref{eq:3x3}) was represented by seven binary variables, so each quantum calculation used 21 logical qubits.
Using somewhat fewer binary variables also gives accurate results, but the D-Wave machine
has many qubits available and Fig.~\ref{fig:3x3} confirms that results remain robust even when this larger number of interconnected qubits is used.
Our two tunable parameters are $\lambda$ and the chain strength, and approximate tuning is sufficient for each of them.
The optimal range for $\lambda$ is typically slightly above the eigenvalue itself so, when calculating at the sequence of $x$ values shown in
Fig.~\ref{fig:3x3}(a), one can use the eigenvalue obtained at one $x$ as the initial estimate for $\lambda$ at the neighboring $x$.
The chain strength is a positive real value; for this figure we simply used one of the values 1.0, 2.0, 3.0 or 4.0
at each $x$ location.

Each of the nine quantum annealing calculations in Fig.~\ref{fig:3x3}(a) used $1000$ ``reads'' (i.e.\ $1000$ annealing cycles) and
the graph shows the smallest numerical result from each set of $1000$ reads, since the smallest is always the best estimator in a variational approach.
Each read used 20 microseconds of computing time on the quantum annealer.

Figure~\ref{fig:3x3}(b) provides a histogram for the case of $x=0.5$ where the chain strength was set to 2.0 and 10 of the 1000
reads had broken chains.
The histogram contains the 990 unbroken cases.
Because the peak of the distribution is in the bin closest to the correct energy, we would be confident of obtaining an accurate result from
this quantum annealing calculation even if the classical answer had not been available.

The method used here is immediately applicable to larger physics systems (and we will do so momentarily) because
each nonzero entry in any Hamiltonian matrix can be provided directly to the quantum annealer.
(Zero entries never need to be provided.)
This differs from the approach used in gate-based quantum computing where the Hamiltonian must be expressed as a
(possibly long) sequence of quantum gates acting on a qubit register \cite{Klco:2019evd,Davoudi:2020yln}.
In addition, gate-based implementations typically include an unphysical sector in the Hilbert space that can be much larger than the physical sector \cite{Klco:2019evd,Davoudi:2020yln} whereas our quantum annealing calculations involve only the physical Hilbert space.
These quantum annealing advantages come with the notable cost of requiring many more qubits than are needed by gate-based hardware.

\begin{figure}
\includegraphics[width=86mm]{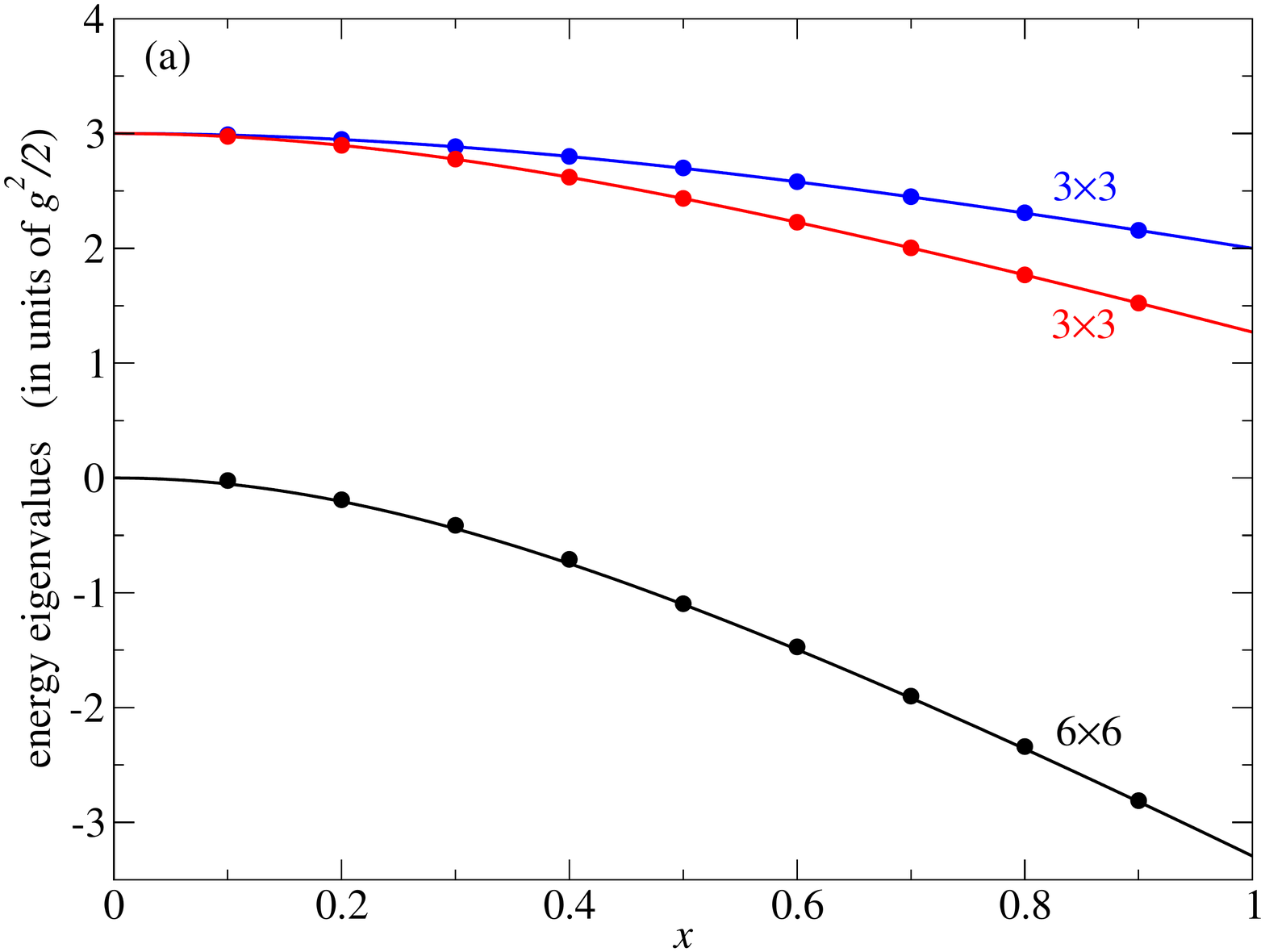}
\includegraphics[width=86mm]{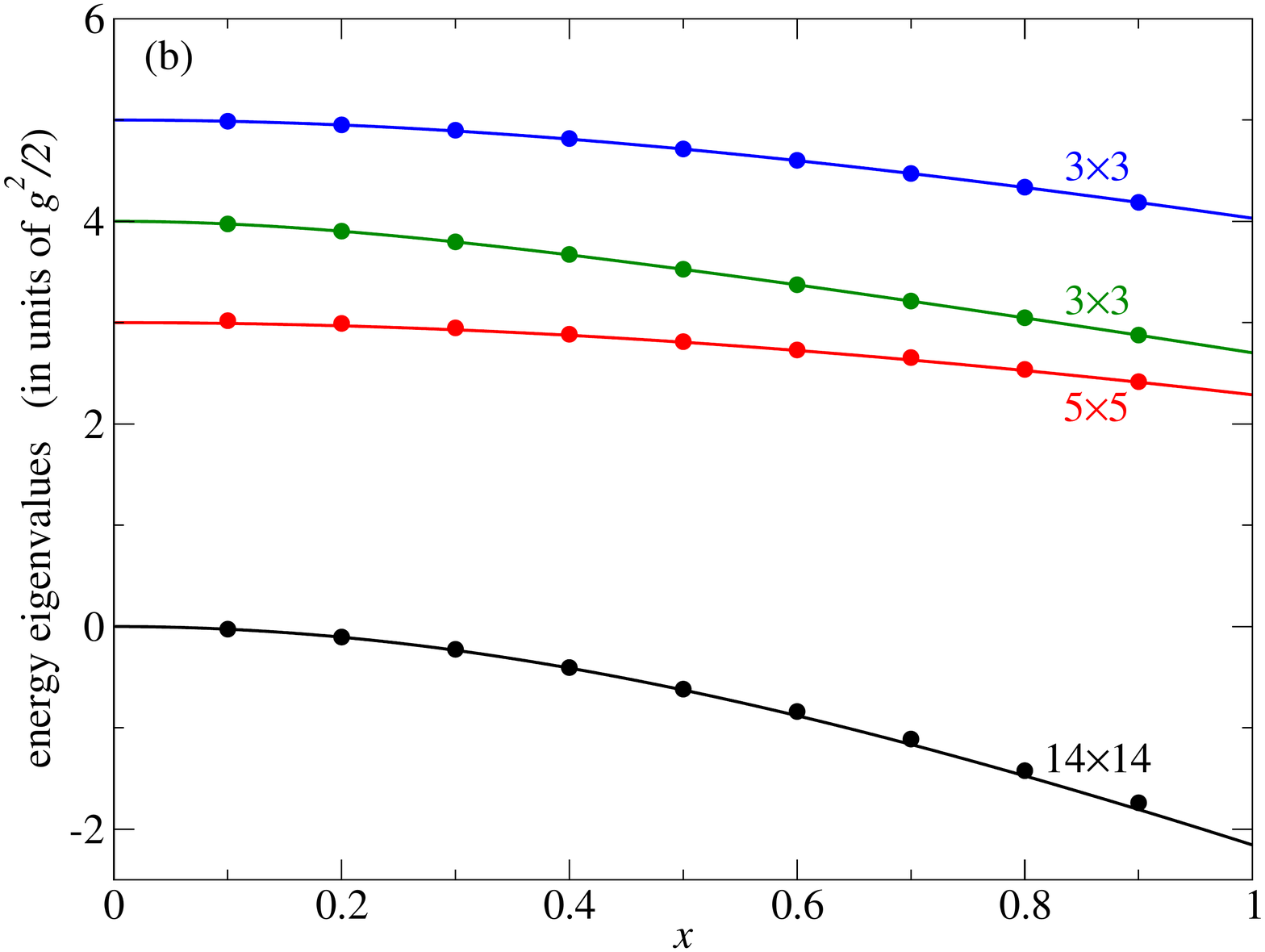}
\caption{(a) Data points show the smallest eigenvalue for each block of a four-plaquette lattice with $j_{\rm max}=\tfrac{1}{2}$,
             computed on the D-Wave Advantage quantum annealer as a function of the gauge coupling $x$ of Eq.~(\ref{eq:defx}).
             Curves are the eigenvalues as computed classically by exact diagonalization.
         (b) As above but for a two-plaquette lattice with $j_{\rm max}=1$.\label{fig:NLO}}
\end{figure}
Figure~\ref{fig:NLO} displays the leading corrections arising from (a) a longer lattice and (b) a larger $j_{\rm max}$.
Specifically, Fig.~\ref{fig:NLO}(a) shows the smallest eigenvalue from each block (determined in Sec.~\ref{sec:symmetries}) for
the four-plaquette lattice with $j_{\rm max}=\tfrac{1}{2}$.
The lowest eigenvalue comes from a 6$\times$6 block and we use $10^4$ reads per $x$ value when running on the quantum annealer,
but the other blocks are 3$\times$3 and 1000 reads will suffice for them.
The number of logical qubits per entry in the state vector is always $K=7$.

Figure~\ref{fig:NLO}(b) shows the smallest eigenvalue from each block of the two-plaquette lattice with $j_{\rm max}=1$, again
choosing $K=7$.
The upper eigenvalues are from a 5$\times$5 block and two 3$\times$3 blocks.
The lowest eigenvalue comes from a 14$\times$14 block where $10^4$ reads is not enough for the original QAE
algorithm \cite{Teplukhin_2019,TeplukhinNature,TeplukhinPCCP,teplukhin2021sampling} with $K=7$.
To continue using a maximum of $10^4$ reads per calculation, we have developed an adaptive version of the QAE algorithm (we call it the AQAE algorithm)
which runs first with $K=4$ to find an approximate solution, then refines the solution by using $K=4$ on a finer grid
in the vicinity of the approximate solution, then zooms in a second time, and then a third time.
This AQAE algorithm uses only $K=4$ qubits per entry in the state vector but after three adaptive steps it attains the
accuracy of $K=4+3=7$.
Additional zooms are possible until the eigenvalue is no longer improved.
Data points along the lowest curve in Fig.~\ref{fig:NLO}(b) were obtained from the AQAE algorithm with $K=4$ and using
between four and nine zooms per $x$ value.
A distinction between the data points and the exact curve becomes visible on the graph for larger $x$ values,
but even the data point at $x=0.9$ only deviates from the exact curve by 4\%.
Details about the AQAE algorithm are provided in Appendix \ref{sec:AQAE}.

\begin{figure}
\includegraphics[width=86mm]{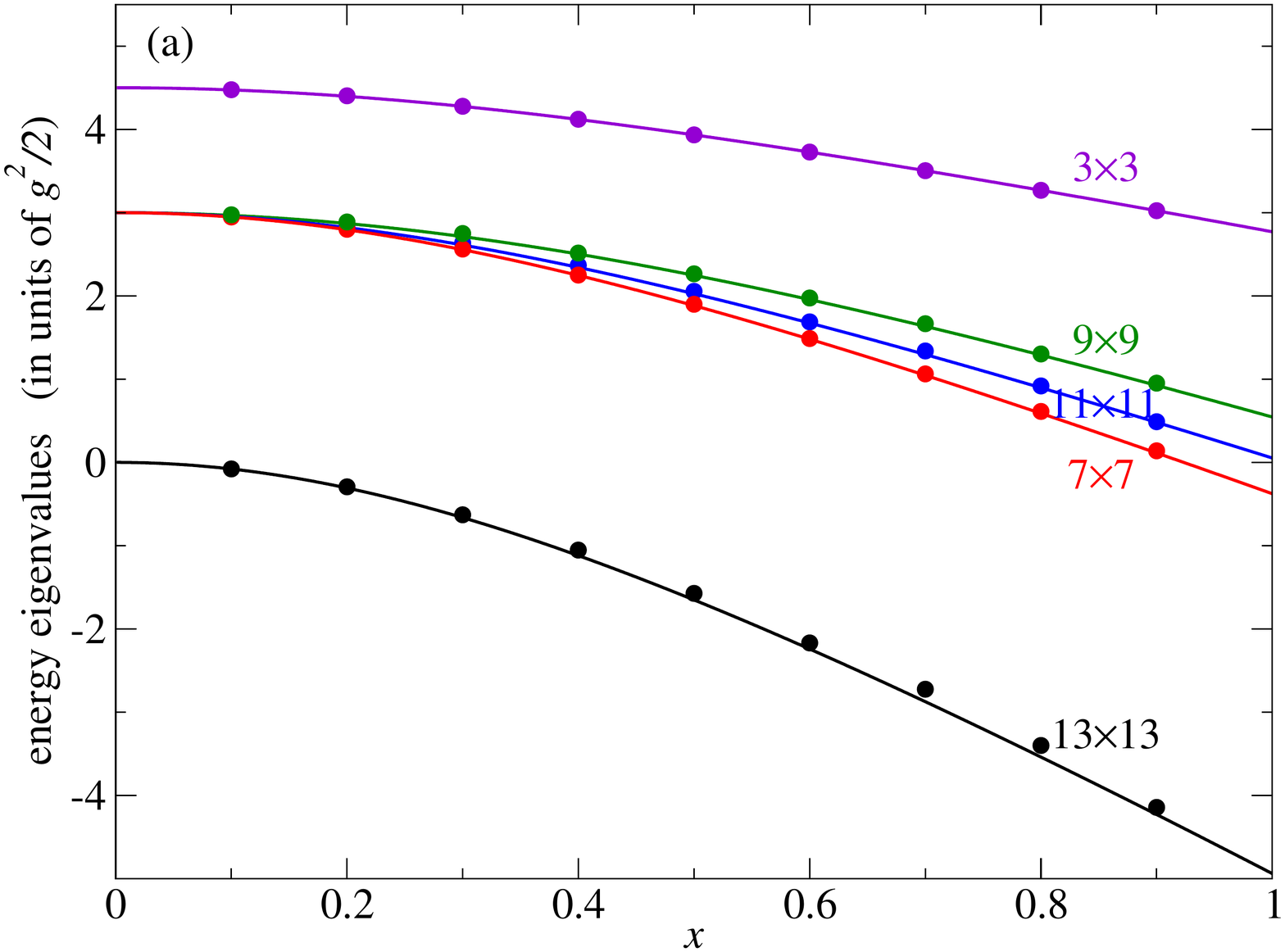}
\includegraphics[width=86mm]{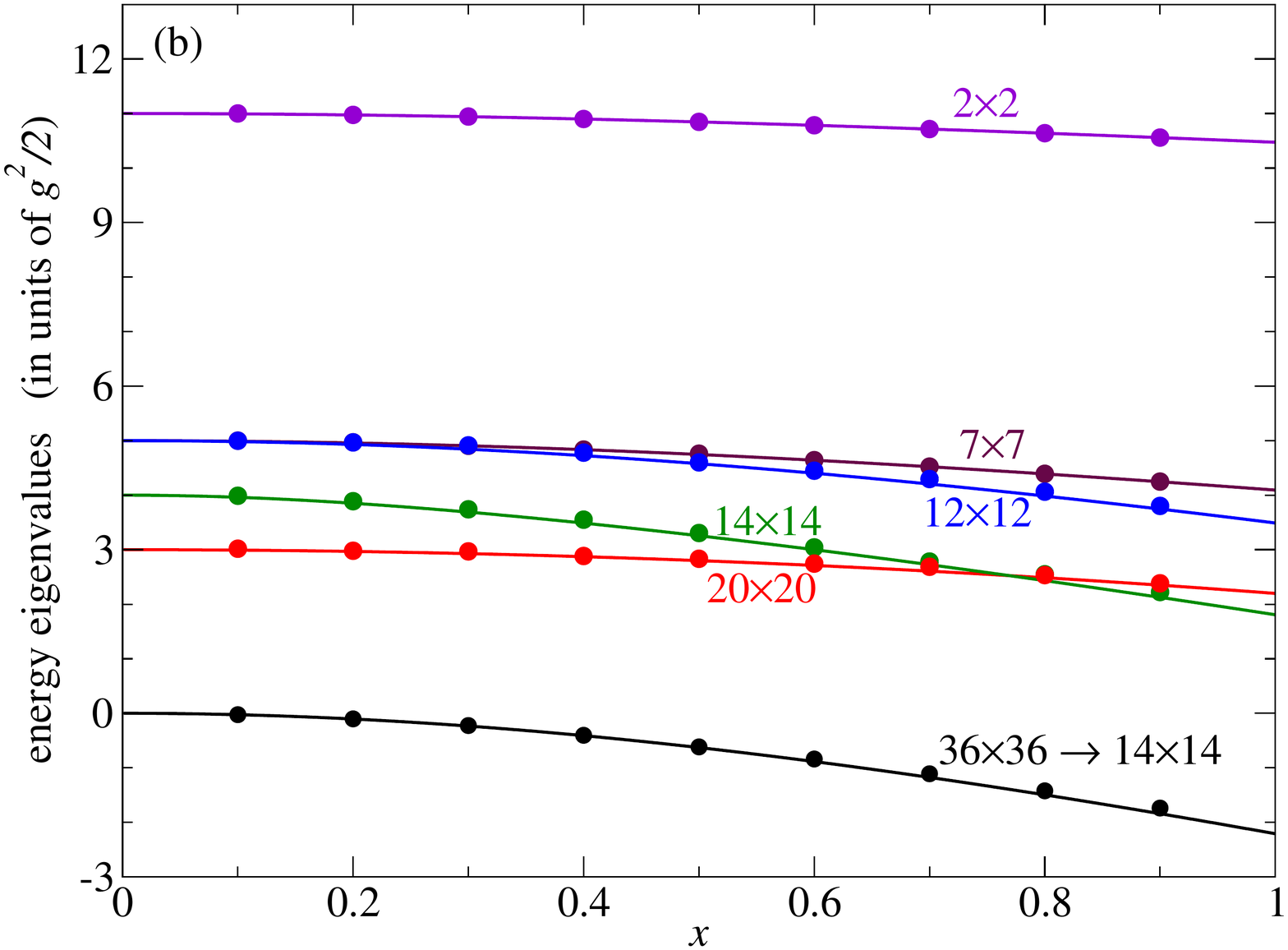}
\caption{(a) Data points show the smallest eigenvalue for each block of a six-plaquette lattice with $j_{\rm max}=\tfrac{1}{2}$,
             computed on the D-Wave Advantage quantum annealer as a function of the gauge coupling $x$ of Eq.~(\ref{eq:defx}).
             Curves are the eigenvalues as computed classically by exact diagonalization.
         (b) As above but for a two-plaquette lattice with $j_{\rm max}=\tfrac{3}{2}$.\label{fig:NNLO}}
\end{figure}
Next-order corrections are obtained by (a) extending the lattice to six plaquettes and (b)
extending $j_{\rm max}$ to $\tfrac{3}{2}$.
Both of these cases can be studied with our same \textsc{ocean} code and results are displayed in Fig.~\ref{fig:NNLO}.
All of the blocks that generate the excited states in Figs.~\ref{fig:NNLO}(a) and \ref{fig:NNLO}(b) are readily handled by our
AQAE algorithm with $10^4$ reads.
The lowest block in Fig.~\ref{fig:NNLO}(a) begins to deviate from the curve as $x$ grows, but using more than
$10^4$ reads would allow that deviation to shrink.
The lowest block in Fig.~\ref{fig:NNLO}(b) is 36$\times$36, and $10^4$ reads are insufficient to see any
significant improvement beyond the 14$\times$14 results that were already shown in Fig.~\ref{fig:NLO}(b).
Recalling that the 36$\times$36 block contains this exact 14$\times$14 matrix, we can conclude that the extra 22 basis states
make negligible contributions at the resolution of Fig.~\ref{fig:NNLO}(b).
Therefore we show the same results from the 14$\times$14 truncation of the ground state in Figs.~\ref{fig:NLO}(b) and
\ref{fig:NNLO}(b).

To confirm that our AQAE \textsc{ocean} code is working correctly, the same code was written to run on
either the D-Wave quantum annealer or on a laptop with classical simulated annealing by simply
changing a single integer flag in the code.
We have verified that the output from classical simulated annealing is in excellent agreement with all exact curves,
including the full 36$\times$36 matrix for Fig.~\ref{fig:NNLO}(b).

\begin{figure}
\includegraphics[width=86mm]{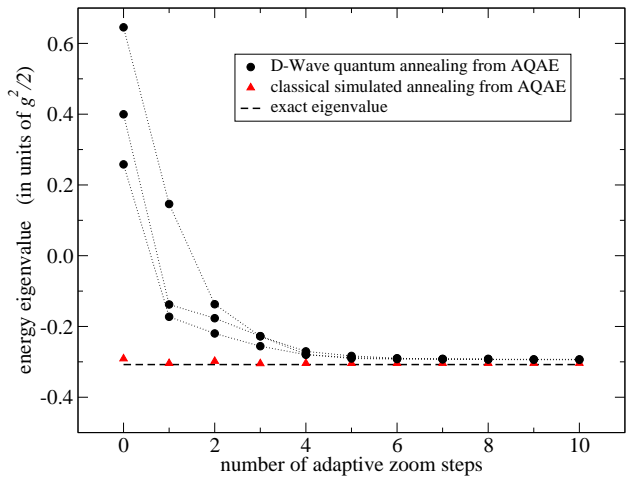}
\caption{Three runs of the AQAE algorithm at each adaptive step using $K=4$ logical qubits per entry in the state vector.
         This example is the ground state of a six-plaquette lattice with $j_{\rm max}=\tfrac{1}{2}$ at $x=0.2$.
         The use of several adaptive steps is visibly crucial on D-Wave hardware but less important for the noise-free case of classical
         simulated annealing.\label{fig:adaptive}}
\end{figure}

Figure~\ref{fig:adaptive} offers an example of how the AQAE algorithm has been vital to the results obtained in this work.
Without any adaptive steps, the energy eigenvalue in that graph is clearly far above the correct result when running on D-Wave hardware.
There would also be a large error bar as seen by comparing the three separate calculations (each from $10^4$ reads)
displayed in Fig.~\ref{fig:adaptive}.
The first adaptive step provides a major improvement and successive steps continue to approach the true result,
thus allowing the completion of Figs.~\ref{fig:NLO} and \ref{fig:NNLO} where statistical error bars are smaller
than the data symbols.
In contrast, the adaptive steps provide smaller improvements when calculating with a classical simulator.
Figure~\ref{fig:adaptive} shows that classical simulated annealing from the original QAE algorithm
(corresponding to no adaptive zoom steps) is already within about 5\% of the correct result.

\section{Computing vacuum expectation values\label{sec:vevs}}

The particles contained within SU(2) pure gauge theory are called glueballs, and their energies are obtained from differences
between the eigenvalues calculated in Sec.~\ref{sec:eigenvalues}, $E_i-E_0$, where $E_0$ is the smallest eigenvalue.
The symmetries implemented in Sec.~\ref{sec:symmetries} identify the specific parity and momentum corresponding to each $E_i$.
Physically interesting glueball energies would be obtained from computations on larger lattices closer to the continuum limit,
which is approached as the inverse gauge coupling $x$ is increased.

The calculations in Sec.~\ref{sec:eigenvalues} produced eigenvectors as well as eigenvalues, and the eigenvector corresponding to $E_0$
represents the theory's vacuum state.
This provides access to the calculation of the vacuum-to-vacuum matrix elements that are so important in quantum field theory.
In this section, vacuum expectation values are computed and used to probe the systematic effects due to
lattice volume and the $j_{\rm max}$ truncation.

Because we cannot use an infinite number of qubits, there is always some limit to the precision of any calculation.
For the variational method, these uncertainties are $O(\epsilon)$ for the eigenstates but $O(\epsilon^2)$ for the eigenvalues,
where $\epsilon$ represents a perturbation.
Therefore we can anticipate less precise results for vacuum expectation values than for the associated eigenvalue.

\begin{figure}
\includegraphics[width=86mm]{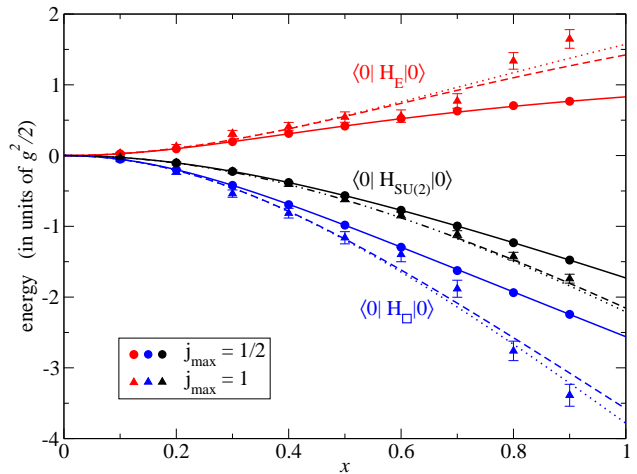}
\caption{The chromoelectric part, plaquette part, and total vacuum expectation value for the SU(2) Hamiltonian on a two-plaquette
         lattice
         as functions of the gauge coupling $x$ of Eq.~(\ref{eq:defx}).  Data points are computed on the D-Wave Advantage quantum annealer.
         Solid, dashed and dotted curves are classical calculations for $j_{\rm max}=\tfrac{1}{2}$, 1 and $\tfrac{3}{2}$
         respectively.\label{fig:Hvevjmax}}
\end{figure}
Recall from Eq.~(\ref{eq:H}) that the general SU(2) Hamiltonian is the sum of a chromoelectric term and a plaquette term,
where the continuum limit of the plaquette term contains a chromomagnetic contribution and an additive constant.
In units of $g^2/2$, the vacuum expectation value of Eq.~(\ref{eq:H}) can be written as
\begin{equation}\label{eq:vevs}
\left<0\right|H_{\rm SU(2)}\left|0\right> = \left<0\right|H_E\left|0\right> + \left<0\right|H_\square\left|0\right> \,.
\end{equation}
To the left of the equal sign is the smallest eigenvalue.
The first(second) term on the right side can be calculated by matrix multiplication using the diagonal(off-diagonal)
terms in the Hamiltonian together with the ground-state eigenvector that was computed in Sec.~\ref{sec:eigenvalues}.

Figure~\ref{fig:Hvevjmax} shows the three terms of Eq.~(\ref{eq:vevs}) on a two-plaquette lattice for the available $j_{\rm max}$ values, with data points obtained from the quantum annealer and curves obtained classically.
Data points show small but visible deviations from the classical curves for the chromoelectric and
plaquette terms separately but, as anticipated, their sum is equal to the minimum eigenvalue and is closer to its classical curve.
Data points for $j_{\rm max}=\tfrac{3}{2}$ do not appear on the graph because, as discussed in Sec.~\ref{sec:eigenvalues},
those D-Wave results are not significantly resolved from the $j_{\rm max}=1$ data points.

The full effects of gauge fields are attained as $j_{\rm max}\to\infty$, and
the comparison of different $j_{\rm max}$ choices in Fig.~\ref{fig:Hvevjmax} suggests a rapid convergence for the range of
gauge couplings studied here, $0<x<1$.
The precise rate of convergence always depends on the particular observable being considered, and we see that
calculations for $j_{\rm max}=1$ and $\tfrac{3}{2}$ are closer together for the full Hamiltonian
than for the chromoelectric or plaquette terms separately.

\begin{figure}
\includegraphics[width=86mm]{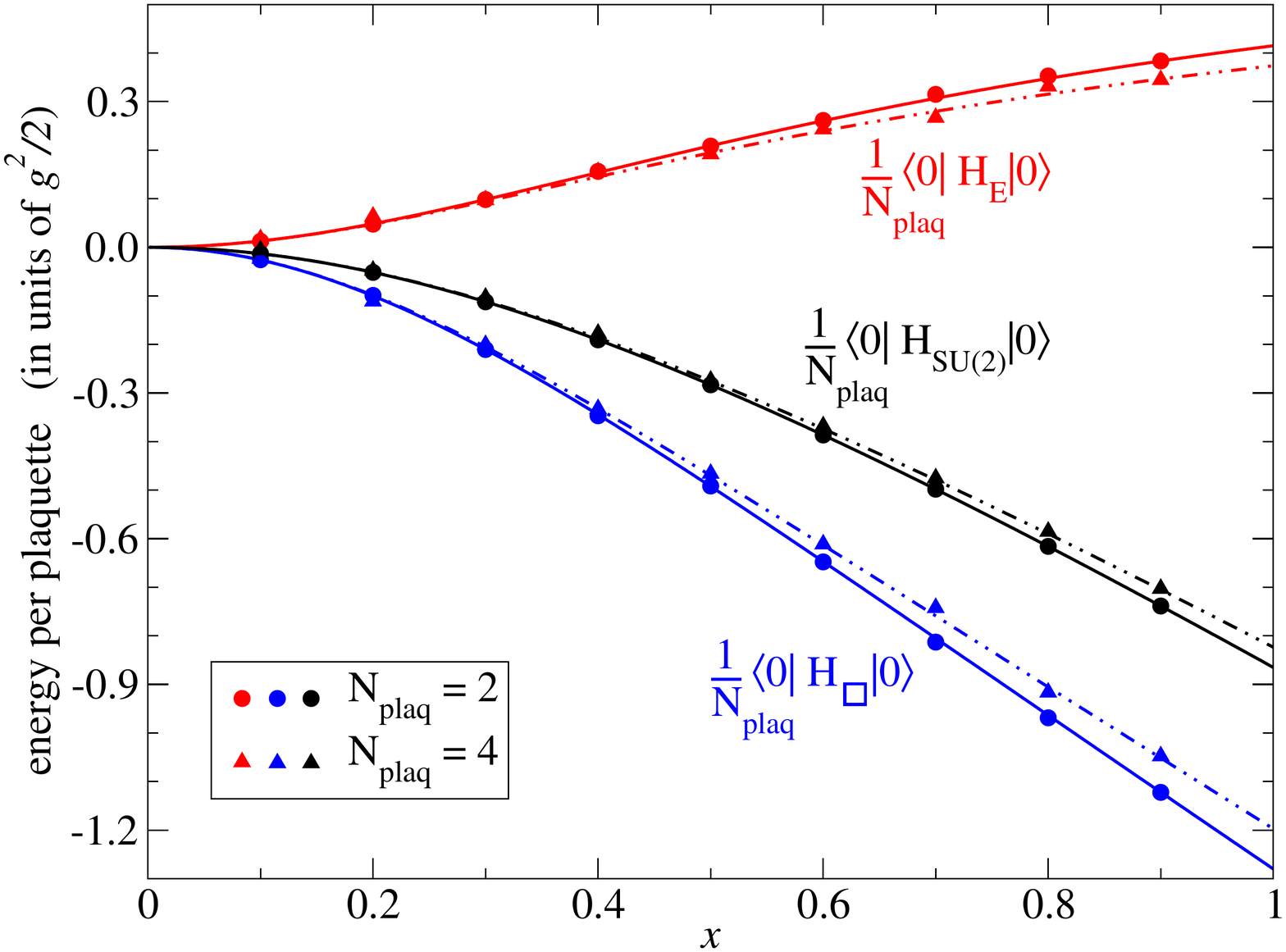}
\caption{The chromoelectric part, plaquette part, and total vacuum expectation value for the SU(2) Hamiltonian with
         $j_{\rm max}=\tfrac{1}{2}$
         as functions of the gauge coupling $x$ of Eq.~(\ref{eq:defx}).  Data points are computed on the D-Wave Advantage quantum annealer.
         Solid, dashed and dotted curves are classical calculations for $N_{\rm plaq}=2$, $4$, and $6$
         respectively.\label{fig:Hvevvolume}}
\end{figure}
To determine how results depend on lattice volume, it is convenient to divide Eq.~(\ref{eq:vevs}) by the number of plaquettes,
thereby obtaining an energy density.
Classical calculations show no visible distinction between $N_{\rm plaq}=4$ and 6 at the resolution of Fig.~\ref{fig:Hvevvolume}
so they appear as a single dot-dashed curve.
The two-plaquette result is a nearby solid curve.
Taken together, the three volumes show that these energy densities (chromoelectric, plaquette, and total) are indeed local
quantities with no significant dependence on lattice volume beyond a few plaquettes, at least for the range of $x$ considered
here.
As expected, computations on the D-Wave quantum annealer in Fig.~\ref{fig:Hvevvolume}
show smaller errors for the total energy than for the separate chromoelectric and plaquette terms.

\section{Computing time evolution\label{sec:time}}

Time evolution is a key motivation for using the Hamiltonian approach because traditional lattice gauge theory calculations employ
Euclidean time and thus lack access to real-time dynamics.
For quantum computing, real-time evolution can be handled with the Suzuki-Trotter
approach \cite{Trotter,Suzuki}, where $e^{-i\epsilon H}$ is applied repeatedly for a small time step $\epsilon$,
and $H$ denotes the Hamiltonian operator.
Since a quantum annealer does not provide gates from which to build an operator, time evolution must instead
be translated into a ground-state eigenvalue problem.
This can be accomplished by using Kitaev-Feynman clock states \cite{Feynman:85,Kitaev} that were also used, for example,
to show the equivalence of adiabatic quantum computing to gate-based quantum computing \cite{1366223,Aharonov}.

The basic idea is that a sequence of time values is defined, and the quantum annealer will calculate the minimum eigenvalue
for the entire time sequence at once, thereby giving the state of the system at all times.
A clear derivation can be found in Ref.~\cite{McCleanE3901} where the approach is called the time-embedded discrete variational principle (TEDVP).
Reference \cite{McCleanE3901} shows that the functional to be minimized is
\begin{equation}\label{eq:timefunctional}
{\cal L} = \sum_{t,t^\prime}\left<t^\prime\right|\left<\Psi_{t^\prime}\right|{\cal C}\left|\Psi_t\right>\left|t\right>
         - \lambda\left(\sum_{t,t^\prime}\left<t^\prime\right|\left<\Psi_{t^\prime}|\Psi_t\right>\left|t\right>-1\right)
\end{equation}
where $\left|\Psi_t\right>$ is the state of the system at time $t$, $\left|t\right>$ is the state of the clock at time $t$,
and the clock Hamiltonian is
\begin{eqnarray}
{\cal C} &=& C_0 + \frac{1}{2}\sum_t\bigg(I\otimes\left|t\right>\left<t\right|-U_t\otimes\left|t+\epsilon\right>\left<t\right| \nonumber \\
         & & -U_t^\dagger\otimes\left|t\right>\left<t+\epsilon\right|+I\otimes\left|t+\epsilon\right>\left<t+\epsilon\right|\bigg)
\label{eq:clockH}
\end{eqnarray}
where $U_t=e^{-i\epsilon H_t}$ performs the evolution from $t$ to $t+\epsilon$ and $C_0$ is a penalty term used to specify the initial
state at time $t=0$.
Because Eq.~(\ref{eq:timefunctional}) has the same form as Eq.~(\ref{eq:F}) up to an additive constant, we can apply the QAE
directly to the TEDVP calculation of time evolution.
Notice that the clock Hamiltonian acts on a compound state built from both the physical state and the clock state.
The compound state is larger than the physical state by a factor of the number of time steps, so the
number of qubits required will increase by this same multiplicative factor.

Implementing a QAE+TEDVP algorithm on the quantum annealer is hampered by imaginary terms in $U_t$ because
the D-Wave hardware needs real entries throughout Eq.~(\ref{eq:ising}).
We will handle this by working in a basis where the Hamiltonian is purely imaginary.

Consider the case of a two-plaquette lattice that is described sufficiently accurately by the truncated Hamiltonian shown in
Eq.~(\ref{eq:3x3}).
The 3$\times$3 block describes the ground state plus two excited states, and we want to calculate the oscillation between
those two excited states as a function of time.
As indicated by the labels in Eq.~(\ref{eq:3x3}), this will be oscillation between a superposition of single-plaquette excitations
$\tfrac{1}{\sqrt{2}}\left(\left|2_2^22_1^1\right>+\left|2_1^12_2^2\right>\right)$ and a pair of round-the-world excitations
$\left|1_2^21_2^2\right>$.
Both options have exactly four excited gauge links, so they are exactly degenerate at $x=0$ and nearly degenerate for small $x$.

There is no change of basis that converts this 3$\times$3 block into a 3$\times$3 imaginary matrix.
However, in the strong coupling (small $x$) region where Eq.~(\ref{eq:3x3}) applies, we can augment it with an additional heavy state
to form this matrix,
\begin{equation}\label{eq:heavystate}
H_{\rm new} = \frac{g^2}{2}\left(\begin{array}{cccc} 0 & -2\sqrt{2}x & 0 & 0 \\ -2\sqrt{2}x & 3 & -\tfrac{x}{\sqrt{2}} & 0 \\
              0 & -\tfrac{x}{\sqrt{2}} & 3 & -2\sqrt{2}x \\ 0 & 0 & -2\sqrt{2}x & 6 \end{array}\right)
\end{equation}
that can be written (up to a constant) in a purely imaginary form,
\begin{equation}\label{eq:PHP}
P^{-1}H_{\rm new}P = \frac{g^2}{2}\left(\begin{array}{cccc} 3 & 0 & 0 & -ih_- \\ 0 & 3 & -ih_+ & 0 \\
                     0 & ih_+ & 3 & 0 \\ ih_- & 0 & 0 & 3 \end{array}\right) \,,
\end{equation}
\begin{equation}
h_\pm = \frac{1}{2}\sqrt{18+33x^2\pm\sqrt{65x^4+1116x^2+324}} \,.
\end{equation}
Since our calculation of oscillations will only depend on energy differences, the constant can be
subtracted from Eq.~(\ref{eq:PHP}), leaving an imaginary matrix that is block diagonal.
The block with $h_+$ contains the ground state and the fictitious heavy state.
The block with $h_-$ contains the two intermediate states that are of interest to us, and its time evolution is represented by
\begin{equation}\label{eq:oscillations}
U_t = \left(\begin{array}{rr} \cos(\omega\epsilon) & -\sin(\omega\epsilon) \\
      \sin(\omega\epsilon) & \cos(\omega\epsilon) \end{array}\right) \,,
\end{equation}
where $\omega=h_-g^2/2$.

\begin{figure}
\includegraphics[width=80mm]{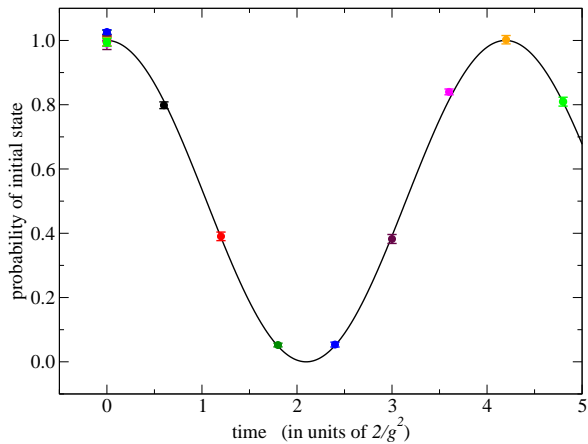}
\caption{Probability oscillations between the two states of Eq.~(\ref{eq:oscillations}).
         The data points were computed on the D-Wave Advantage quantum annealer at gauge coupling $x=0.1$ using two time steps
         per run ($t=0$ and one $t\neq0$).
         Each run has its own color at two locations on the graph.
         The exact curve is shown for comparison: $\cos^2(\omega t)$ with $\omega=h_-g^2/2$.\label{fig:oscillateK7}}
\end{figure}
Note that, for small $x$, the effect of adding the fictitious heavy state [which has the value 6 on the diagonal of Eq.~(\ref{eq:heavystate})]
is smaller than the effect of truncating the Hamiltonian matrix down to its 3$\times$3 form
[because Eq.~(\ref{eq:N2j1}) has values smaller than 6 on the diagonal that must be truncated to arrive at the 3$\times$3 form].
Therefore, since we are considering an example where the 3$\times$3 truncation is sufficiently accurate, we are justified to add the
heavy state.

It is now straightforward to implement our QAE+TEDVP code on D-Wave hardware.
However, today's quantum annealers are only intended to handle situations that do not have sign problems.
Specifically, D-Wave hardware is designed to handle ``stoquastic'' Hamiltonian matrices, which have only nonpositive off-diagonal elements
in the computational basis \cite{McGeoch}, but Eq.~(\ref{eq:clockH}) does not have this form when $U_t$ is given by Eq.~(\ref{eq:oscillations}) with an arbitrary time step $\epsilon$.
Figure \ref{fig:oscillateK7} presents the output from several runs of our code that used only two times each: $t=0$ and a larger $t=\epsilon$.
With two states at two times and $K=7$, each computation used 28 logical qubits.
To obtain the precision of Fig.~\ref{fig:oscillateK7}, we used $5\times10^4$ reads.
However, even noisy data would be sufficient
to provide a useful estimate of the
frequency $\omega$ directly from the D-Wave data.
This is valuable because we can then choose the truly stoquastic case of $\epsilon=\pi/\omega$ for followup computations,
where D-Wave hardware will handle time evolution well.
\begin{figure}
\includegraphics[width=80mm]{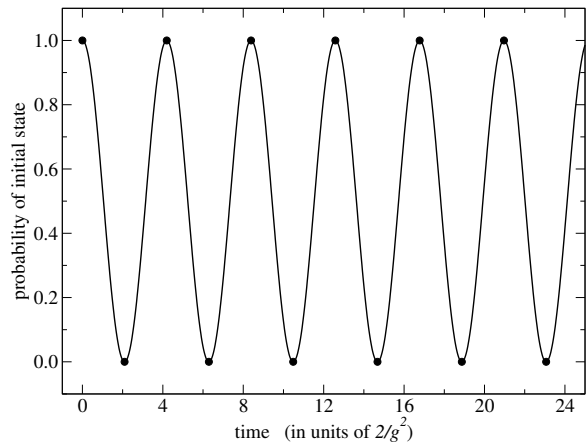}
\caption{Probability oscillations between the two states of Eq.~(\ref{eq:oscillations}).
         The data points were computed on the D-Wave Advantage quantum annealer at gauge coupling $x=0.1$.
         All 12 points are from a single run that used $10^4$ reads with $\lambda=0.12$ and the chain strength set to 0.3.
         The exact curve is shown for comparison: $\cos^2(\omega t)$ with $\omega=h_-g^2/2$.\label{fig:oscillateK2}}
\end{figure}

Figure~\ref{fig:oscillateK2} shows 12 time values obtained from a single run of stoquastic time evolution using our QAE+TEDVP
algorithm on the D-Wave Advantage hardware.
This computation used $K=2$ for each of the two states at each of 12 time values for a total of 48 logical qubits.
The physical energy gap between the two oscillating states can be extracted as $E_2-E_1=2\omega$.

For an indication of how D-Wave hardware performance has been improving, we compare our findings with Fig.~1 of
Ref.~\cite{jalowiecki}.
In that work, a similar method was used to observe Rabi oscillations by using the previous two generations of D-Wave annealers,
the original 2000Q (released in 2017) and the low-noise 2000Q (released in 2019).
The authors of Ref.~\cite{jalowiecki} used those machines to attain five time values and six time values respectively in their
Rabi oscillation computations.  Our study has used the newer Advantage hardware (released in 2020)
to attain 12 time values in Fig.~\ref{fig:oscillateK2}.
\section{Summary and outlook\label{sec:summary}}

In this work, we have used a D-Wave quantum annealer to compute several quantities in a non-Abelian gauge theory.
Although the annealer is designed to focus on the optimization problem for an Ising model, we have demonstrated computations in SU(2) lattice gauge theory for eigenvalues, eigenvectors, vacuum expectation values, and real-time evolution.
Quantum computing is presently in an era of noisy qubits, but the graphs in Figs.~\ref{fig:3x3}-\ref{fig:oscillateK2} demonstrate that a quantum annealer can already produce results that are reasonably precise and accurate, at least on lattices having only a few plaquettes.

For the one-dimensional plaquette lattices studied in this work, classical preprocessing readily produced the explicit Hamiltonian matrices, and then translation and reflection symmetries were used to block diagonalize those matrices.
The QAE algorithm \cite{Teplukhin_2019,TeplukhinNature,TeplukhinPCCP,teplukhin2021sampling} allowed each of those blocks to be entered directly into the quantum annealer without the need to construct them from products of quantum gates.
The D-Wave hardware readily determined respectable results for the lowest eigenvalue and eigenvector from each of the smaller blocks.
To extend this success to larger blocks, we developed an adaptive QAE algorithm which is presented in Appendix \ref{sec:AQAE}.
The D-Wave \textsc{ocean} software \cite{Ocean} offers several methods for tuning hardware performance and enhancing results, but we obtained good quality output without these aids, retaining only the two mandatory adjustable quantities: the penalty parameter $\lambda$ of QAE and the chain strength of D-Wave.
Each of our calculations needed only a portion of the physical Hilbert space, and no unphysical Hilbert space was present.

Real-time oscillations between two excited states were also computed on the quantum annealer.
The basic approach was to use Kitaev-Feynman clock states \cite{Feynman:85,Kitaev} and we implemented these by combining the QAE with the TEDVP \cite{McCleanE3901}.
Besides demonstrating the ability to access time evolution on a quantum annealer, our QAE+TEDVP computation is also a method for measuring the energy splitting between excited states.

Our choice to begin with SU(2) pure gauge theory on a two-plaquette lattice follows the work of Ref.~\cite{Klco:2019evd}, where the same system was studied on a gate-based quantum computer.
Our work goes beyond that starting point in two ways: by extending the number of plaquettes from $N_{\rm plaq}=2$ to 4 and 6, and by increasing the gauge field truncation from $j_{\rm max}=\tfrac{1}{2}$ to 1 and $\tfrac{3}{2}$.
For the range of gauge couplings in our work, calculations at these sequences of $N_{\rm plaq}$ and $j_{\rm max}$ values indicate that contributions which would have arisen from still larger $N_{\rm plaq}$ and $j_{\max}$ values are negligible.
All of our Hamiltonians were straightforward to implement on the quantum annealer because each new matrix can be encoded directly, without any decomposition into gates, but this advantage comes at the expense of requiring many more qubits than the gate-based approach of Ref.~\cite{Klco:2019evd}.

Gate-based quantum computers have the ability to store a quantum state more efficiently than classical computers, and this will be very significant for lattices that are large enough to address the intended goals of nuclear and particle physics phenomenology.
Indeed, this state-storage scaling advantage is a key motivator for Ref.~\cite{Klco:2019evd} and for the entire field of quantum computation in lattice gauge theories.
If quantum annealers are to compete with gate-based hardware, then the storage scaling issue must be addressed by quantum annealers as they continue to evolve toward a fully universal form of adiabatic quantum computing. That will be an important challenge for future algorithms on future adiabatic hardware but, independently, near-term quantum annealers might be useful alongside gate-based hardware in complementary rather than competitive ways.
For example, some algorithms relevant to lattice gauge theory might achieve a speedup sufficient to supersede classical computers
\cite{King,PhysRevX.8.031016,PhysRevA.65.042308,PhysRevLett.109.050501}, making quantum annealers a useful tool during the next several years of study on intermediate-sized lattices. To assess the power of quantum annealing, further experiments are called for \cite{Preskill_2018}, and our work is a step along this path.
As quantum hardware continues to evolve rapidly,
we anticipate that mutually complementary roles could emerge for several quantum and classical hardware platforms within the broad scope of lattice gauge theory research.

\appendix
\begin{widetext}
\section{Deriving the plaquette term}\label{sec:derivemag}

To derive Eq.~(\ref{eq:oneplaq}), begin with the constraint of Gauss's law at lattice site 1 of Fig.~\ref{fig:lattice},
which says the three gauge links form a color singlet, $\left|j,m\right>_1=\left|0,0\right>_1$, giving
\begin{eqnarray}
\left|0,0\right>_1
&=& \sum_{m_I}\left|j_I,-m_I,j_I,m_I\right>_1\left<j_I,-m_I,j_I,m_I|0,0\right>_1 \\
&=& \sum_{m_I}\left|j_I,-m_I,j_I,m_I\right>_1\frac{(-1)^{j_I+m_I}}{\sqrt{2j_I+1}} \\
&=& \sum_{m_I}\frac{(-1)^{j_I+m_I}}{\sqrt{2j_I+1}}\sum_{m_A^\prime}\sum_{m_E}\left|j_A,m_A^\prime,j_E,m_E,j_I,m_I\right>_1
    \left<j_A,m_A^\prime,j_E,m_E,j_I,m_I|j_I,-m_I,j_I,m_I\right>_1 \\
&=& (-1)^{j_A-j_E+j_I}\sum_{m_A^\prime}\sum_{m_E}\sum_{m_I}\left|j_A,m_A^\prime,j_E,m_E,j_I,m_I\right>_1
    \left(\begin{array}{ccc} j_A & j_E & j_I \\ m_A^\prime & m_E & m_I \end{array}\right)
\end{eqnarray}
where our convention is to use $m$ for the left or bottom end of a link and to use $m^\prime$ for the right or top end.
The other seven vertices have similar expressions,
\begin{eqnarray}
\left|0,0\right>_2
&=& (-1)^{j_C-j_E+j_J}\sum_{m_C}\sum_{m_E^\prime}\sum_{m_J}\left|j_C,m_C,j_E,m_E^\prime,j_J,m_J\right>_2
    \left(\begin{array}{ccc} j_C & j_E & j_J \\ m_C & m_E^\prime & m_J \end{array}\right) \,, \\
\left|0,0\right>_3
&=& (-1)^{j_C-j_G+j_K}\sum_{m_C^\prime}\sum_{m_G}\sum_{m_K}\left|j_C,m_C^\prime,j_G,m_G,j_K,m_K\right>_3
    \left(\begin{array}{ccc} j_C & j_G & j_K \\ m_C^\prime & m_G & m_K \end{array}\right) \,, \\
\left|0,0\right>_4
&=& (-1)^{j_B-j_G+j_L}\sum_{m_A}\sum_{m_G^\prime}\sum_{m_L}\left|j_A,m_A,j_G,m_G^\prime,j_L,m_L\right>_4
    \left(\begin{array}{ccc} j_A & j_G & j_L \\ m_A & m_G^\prime & m_L \end{array}\right) \,, \\
\left|0,0\right>_5
&=& (-1)^{j_B-j_F+j_I}\sum_{m_B^\prime}\sum_{m_F}\sum_{m_I^\prime}\left|j_B,m_B^\prime,j_F,m_F,j_I,m_I^\prime\right>_5
    \left(\begin{array}{ccc} j_B & j_F & j_I \\ m_B^\prime & m_F & m_I^\prime \end{array}\right) \,, \\
\left|0,0\right>_6
&=& (-1)^{j_D-j_F+j_J}\sum_{m_D}\sum_{m_F^\prime}\sum_{m_J^\prime}\left|j_D,m_D,j_F,m_F^\prime,j_J,m_J^\prime\right>_6
    \left(\begin{array}{ccc} j_D & j_F & j_J \\ m_D & m_F^\prime & m_J^\prime \end{array}\right) \,, \\
\left|0,0\right>_7
&=& (-1)^{j_D-j_H+j_K}\sum_{m_D^\prime}\sum_{m_H}\sum_{m_K^\prime}\left|j_D,m_D^\prime,j_H,m_H,j_K,m_K^\prime\right>_7
    \left(\begin{array}{ccc} j_D & j_H & j_K \\ m_D^\prime & m_H & m_K^\prime \end{array}\right) \,, \\
\left|0,0\right>_8
&=& (-1)^{j_B-j_H+j_L}\sum_{m_B}\sum_{m_H^\prime}\sum_{m_L^\prime}\left|j_B,m_B,j_H,m_H^\prime,j_L,m_L^\prime\right>_8
    \left(\begin{array}{ccc} j_B & j_H & j_L \\ m_B & m_H^\prime & m_L^\prime \end{array}\right) \,.
\end{eqnarray}
The product of the eight vertex states defines the state of the entire lattice.
Notice that we always list the gauge links $A$ through $L$ in alphabetical order
so the calculation will be self-consistent and have the correct Clebsch-Gordan phases.
The labeling of gauge links chosen in Fig.~\ref{fig:lattice} (even horizontal, then odd horizontal, then vertical) is not required,
but it does maintain a convenient pattern among the four plaquette operators during the derivations.
The first plaquette operator is
\begin{equation}
\square_1 = \sum_{s_1}\sum_{s_2}\sum_{s_6}\sum_{s_5}(-1)^{s_1+s_2+s_6+s_5}U^E_{-s_1,s_2}U^J_{-s_2,s_6}U^F_{s_5,-s_6}U^I_{s_1,-s_5}
\end{equation}
where each sum includes only the two terms $s_i=\pm\tfrac{1}{2}$.
Notice that the subscripts on $U^F$ and $U^I$ have been interchanged because going counterclockwise around the plaquette
means we are going from the $m^\prime$ end to the $m$ end on those two links.
The effect of an operator $U$ is \cite{Byrnes:2005qx,Klco:2019evd}
\begin{eqnarray}
U_{s,s^\prime}\left|j,m,m^\prime\right> &=& \sum_{J=\left|j-\tfrac{1}{2}\right|}^{j+\tfrac{1}{2}}\sqrt{\frac{2j+1}{2J+1}}
\sum_M\sum_{M^\prime}\left<J,M|j,m;\tfrac{1}{2},s\right>\left<J,M^\prime|j,m^\prime;\tfrac{1}{2},s^\prime\right>
\left|J,M,M^\prime\right> \\
&=& \sum_{J=\left|j-\tfrac{1}{2}\right|}^{j+\tfrac{1}{2}}\sqrt{2j+1}\sqrt{2J+1}
   \sum_M\sum_{M^\prime}(-1)^{1-2j+M+M^\prime}\left(\begin{array}{ccc} j & \tfrac{1}{2} & J \\ m & s & -M \end{array}\right)
   \left(\begin{array}{ccc} j & \tfrac{1}{2} & J \\ m^\prime & s^\prime & -M^\prime \end{array}\right)
   \left|J,M,M^\prime\right>~~~~~~~~
\end{eqnarray}
where the sums over $M$ and $M^\prime$ contain only a single nonzero term because the Clebsch-Gordan coefficients
vanish unless $M=m+s$ and $M^\prime=m^\prime+s^\prime$.
Applying $\square_1$ to our initial state gives
\begin{eqnarray}
\square_1\left|\psi_{\rm initial}\right>
&=&\sum_{M_E}\sum_{M_E^\prime}\sum_{M_J}\sum_{M_J^\prime}\sum_{M_F}\sum_{M_F^\prime}\sum_{M_I}\sum_{M_I^\prime}
   \sum_{m_A}\sum_{m_A^\prime}\sum_{m_B}\sum_{m_B^\prime}\sum_{m_C}\sum_{m_C^\prime}\ldots\sum_{m_L}\sum_{m_L^\prime}
   \sum_{J_F}\sum_{J_E}\sum_{J_I}\sum_{J_J}\sum_{s_1}\sum_{s_2}\sum_{s_6}\sum_{s_5} \nonumber \\
&& (-1)^{s_1+s_2+s_6+s_5}
   (-1)^{-2j_E-2j_J-2j_F-2j_I+M_E+M_E^\prime+M_J+M_J^\prime+M_F+M_F^\prime+M_I+M_I^\prime} \nonumber \\
&& \sqrt{2j_E+1}\sqrt{2J_E+1}
   \sqrt{2j_J+1}\sqrt{2J_J+1}
   \sqrt{2j_F+1}\sqrt{2J_F+1}
   \sqrt{2j_I+1}\sqrt{2J_I+1} \nonumber \\
&& \left(\begin{array}{ccc} j_A & j_E & j_I \\ m_A^\prime & m_E & m_I \end{array}\right)
   \left(\begin{array}{ccc} j_C & j_E & j_J \\ m_C & m_E^\prime & m_J \end{array}\right)
   \left(\begin{array}{ccc} j_C & j_G & j_K \\ m_C^\prime & m_G & m_K^\prime \end{array}\right)
   \left(\begin{array}{ccc} j_A & j_G & j_L \\ m_A & m_G^\prime & m_L \end{array}\right) \nonumber \\
&& \left(\begin{array}{ccc} j_B & j_F & j_I \\ m_B^\prime & m_F & m_I^\prime \end{array}\right)
   \left(\begin{array}{ccc} j_D & j_F & j_J \\ m_D & m_F^\prime & m_J^\prime \end{array}\right)
   \left(\begin{array}{ccc} j_D & j_H & j_K \\ m_D^\prime & m_H & m_K^\prime \end{array}\right)
   \left(\begin{array}{ccc} j_B & j_H & j_L \\ m_B & m_H^\prime & m_L^\prime \end{array}\right) \nonumber \\
&& \left(\begin{array}{ccc} j_E & \tfrac{1}{2} & J_E \\ m_E & -s_1 & -M_E \end{array}\right)
   \left(\begin{array}{ccc} j_E & \tfrac{1}{2} & J_E \\ m_E^\prime & s_2 & -M_E^\prime \end{array}\right)
   \left(\begin{array}{ccc} j_J & \tfrac{1}{2} & J_J \\ m_J & -s_2 & -M_J \end{array}\right)
   \left(\begin{array}{ccc} j_J & \tfrac{1}{2} & J_J \\ m_J^\prime & s_6 & -M_J^\prime \end{array}\right) \nonumber \\
&& \left(\begin{array}{ccc} j_F & \tfrac{1}{2} & J_F \\ m_F & s_5 & -M_F \end{array}\right)
   \left(\begin{array}{ccc} j_F & \tfrac{1}{2} & J_F \\ m_F^\prime & -s_6 & -M_F^\prime \end{array}\right)
   \left(\begin{array}{ccc} j_I & \tfrac{1}{2} & J_I \\ m_I & s_1 & -M_I \end{array}\right)
   \left(\begin{array}{ccc} j_I & \tfrac{1}{2} & J_I \\ m_I^\prime & -s_5 & -M_I^\prime \end{array}\right) \nonumber \\
&& \left|j_A,m_A,m_A^\prime\right>\left|j_B,m_B,m_B^\prime\right>\left|j_C,m_C,m_C^\prime\right>\left|j_D,m_D,m_D^\prime\right>
   \left|J_E,M_E,M_E^\prime\right>\left|J_F,M_F,M_F^\prime\right> \nonumber \\
&& \left|j_G,m_G,m_G^\prime\right>\left|j_H,m_H,m_H^\prime\right>
   \left|J_I,M_I,M_I^\prime\right>\left|J_J,M_J,M_J^\prime\right>\left|j_K,m_K,m_K^\prime\right>\left|j_L,m_L,m_L^\prime\right> \,.
\end{eqnarray}
Applying a final state to that result allows all sums to be performed and the answer simplifies to
\begin{eqnarray}
\left<\psi_{\rm final}\right|\square_1\left|\psi_{\rm initial}\right>
&=& (-1)^{j_A+j_B+j_C+j_D+2J_E+2J_F+2j_I+2j_J} \nonumber \\
& & \sqrt{2j_E+1}\sqrt{2J_E+1}\sqrt{2j_J+1}\sqrt{2J_J+1}\sqrt{2j_F+1}\sqrt{2J_F+1}\sqrt{2j_I+1}\sqrt{2J_I+1} \nonumber \\
& & \left\{\begin{array}{ccc} j_A & j_E & j_I \\ \tfrac{1}{2} & J_I & J_E \end{array}\right\}
    \left\{\begin{array}{ccc} j_B & j_F & j_I \\ \tfrac{1}{2} & J_I & J_F \end{array}\right\}
    \left\{\begin{array}{ccc} j_C & j_E & j_J \\ \tfrac{1}{2} & J_J & J_E \end{array}\right\}
    \left\{\begin{array}{ccc} j_D & j_F & j_J \\ \tfrac{1}{2} & J_J & J_F \end{array}\right\}
\end{eqnarray}
as given in Eq.~(\ref{eq:oneplaq}).
Results for $\left<\psi_{\rm final}\right|\square_i\left|\psi_{\rm initial}\right>$ with $i=2,3,4$ can be obtained simply by translation symmetry from the $i=1$ result or by explicit calculation.

The numerical values of the nonzero 6j symbols can be obtained from
\begin{eqnarray}
\left\{\begin{array}{ccc} j_0 & j_1 & j_2 \\ \tfrac{1}{2} & j_2+\tfrac{1}{2} & j_1+\tfrac{1}{2} \end{array}\right\}
&=& (-1)^{1+j_0+j_1+j_2}\sqrt{\frac{(1-j_0+j_1+j_2)(2+j_0+j_1+j_2)}{(2j_1+1)(2j_1+2)(2j_2+1)(2j_2+2)}} \,, \\
\left\{\begin{array}{ccc} j_0 & j_1 & j_2 \\ \tfrac{1}{2} & j_2-\tfrac{1}{2} & j_1+\tfrac{1}{2} \end{array}\right\}
&=& (-1)^{j_0+j_1+j_2}\sqrt{\frac{(1+j_0+j_1-j_2)(j_0-j_1+j_2)}{(2j_1+1)(2j_1+2)2j_2(2j_2+1)}} \,, \\
\left\{\begin{array}{ccc} j_0 & j_1 & j_2 \\ \tfrac{1}{2} & j_2+\tfrac{1}{2} & j_1-\tfrac{1}{2} \end{array}\right\}
&=& (-1)^{j_0+j_1+j_2}\sqrt{\frac{(1+j_0-j_1+j_2)(j_0+j_1-j_2)}{2j_1(2j_1+1)(2j_2+1)(2j_2+2)}} \,, \\
\left\{\begin{array}{ccc} j_0 & j_1 & j_2 \\ \tfrac{1}{2} & j_2-\tfrac{1}{2} & j_1-\tfrac{1}{2} \end{array}\right\}
&=& (-1)^{j_0+j_1+j_2}\sqrt{\frac{(1+j_0+j_1+j_2)(-j_0+j_1+j_2)}{2j_1(2j_1+1)2j_2(2j_2+1)}} \,,
\end{eqnarray}
but a 6j symbol is zero unless all four of its triangle conditions are satisfied \cite{Varshalovich:1988ye,Thompson}.

\section{VACUUM SECTOR MATRICES}\label{sec:vacblocks}

As explained in Sec.~\ref{sec:symmetries}, three spatial symmetries can be used to block diagonalize each Hamiltonian matrix.
One explicit example was provided in Eq.~(\ref{eq:3x3}).
The largest blocks that arise from three additional physics systems are provided here.
Recall that the largest block is always the one containing the vacuum state.

For $N_{\rm plaq}=4$ and $j_{\rm max}=\tfrac{1}{2}$, the vacuum block is
\begin{equation}
H = 
\frac{g^2}{2}\left(
\begin{array}{cccccc}
0 & -4x & 0 & 0 & 0 & 0 \\
-4x & 3 & -2x & 0 & -2\sqrt{2}x & 0 \\
0 & -2x & \tfrac{9}{2} & -2x & 0 & 0 \\
0 & 0 & -2x & 6 & -\tfrac{x}{\sqrt{2}} & -x \\
0 & -2\sqrt{2}x & 0 & -\tfrac{x}{\sqrt{2}} & 6 & 0 \\
0 & 0 & 0 & -x & 0 & 6
\end{array}
\right)
\begin{array}{c}
\left|1_1^11_1^11_1^11_1^1\right> \\
\tfrac{1}{2}(\left|2_2^22_1^11_1^11_1^1\right>+\left|1_1^12_2^22_1^11_1^1\right>+\left|1_1^11_1^12_2^22_1^1\right>+\left|2_1^11_1^11_1^12_2^2\right>) \\
\tfrac{1}{2}(\left|2_1^11_1^12_2^21_2^2\right>+\left|1_2^22_1^11_1^12_2^2\right>+\left|2_2^21_2^22_1^11_1^1\right>+\left|1_1^12_2^21_2^22_1^1\right>) \\
\tfrac{1}{2}(\left|1_2^21_2^22_1^12_2^2\right>+\left|2_2^21_2^21_2^22_1^1\right>+\left|2_1^12_2^21_2^21_2^2\right>+\left|1_2^22_1^12_2^21_2^2\right>) \\
\tfrac{1}{\sqrt{2}}(\left|2_1^12_2^22_1^12_2^2\right>+\left|2_2^22_1^12_2^22_1^1\right>) \\
\left|1_2^21_2^21_2^21_2^2\right>
\end{array}
\end{equation}
with the basis states shown beside their corresponding rows.
For $N_{\rm plaq}=6$ and $j_{\rm max}=\tfrac{1}{2}$, the vacuum block is
\begin{equation}
H = \frac{g^2}{2}\left(
\begin{array}{ccccccccccccc}
0 & -2\sqrt{6}x & 0 & 0 & 0 & 0 & 0 & 0 & 0 & 0 & 0 & 0 & 0 \\
\!\!\!-2\sqrt{6}x\!\!\!\! & 3 & -2x & -4x & \!\!-2\sqrt{2}x & 0 & 0 & 0 & 0 & 0 & 0 & 0 & 0 \\
0 & -2x & \tfrac{9}{2} & 0 & 0 & -2x & -2\sqrt{2}x & 0 & 0 & 0 & 0 & 0 & 0 \\
0 & -4x & 0 & 6 & 0 & -\tfrac{x}{2} & -\sqrt{2}x & -2\sqrt{3}x & 0 & 0 & 0 & 0 & 0 \\
0 & -2\sqrt{2}x & 0 & 0 & 6 & 0 & -2x & 0 & 0 & 0 & 0 & 0 & 0 \\
0 & 0 & -2x & -\tfrac{x}{2} & 0 & 6 & 0 & 0 & -2x & -2x & 0 & 0 & 0 \\
0 & 0 & \!\!\!-2\sqrt{2}x & -\sqrt{2}x & -2x & 0 & \tfrac{15}{2} & 0 & -\tfrac{x}{\sqrt{2}} & -\sqrt{2}x & -2x & 0 & 0 \\
0 & 0 & 0 & -2\sqrt{3}x & 0 & 0 & 0 & 9 & 0 & -\tfrac{\sqrt{3}}{2}x & 0 & 0 & 0 \\
0 & 0 & 0 & 0 & 0 & -2x & -\tfrac{x}{\sqrt{2}} & 0 & \tfrac{15}{2} & 0 & 0 & -2x & 0 \\
0 & 0 & 0 & 0 & 0 & -2x & -\sqrt{2}x & -\tfrac{\sqrt{3}}{2}x & 0 & 9 & 0 & -x & 0 \\
0 & 0 & 0 & 0 & 0 & 0 & -2x & 0 & 0 & 0 & 9 & -\tfrac{x}{\sqrt{2}} & 0 \\
0 & 0 & 0 & 0 & 0 & 0 & 0 & 0 & -2x & -x & -\tfrac{x}{\sqrt{2}} & 9 & -\frac{\sqrt{3}}{\sqrt{2}}x\!\!\! \\
0 & 0 & 0 & 0 & 0 & 0 & 0 & 0 & 0 & 0 & 0 & -\frac{\sqrt{3}}{\sqrt{2}}x & 9
\end{array}
\right)
\begin{array}{c}
\left|P_1^{(0)}\right> \\
\left|P_2^{(0)}\right> \\
\left|P_3^{(0)}\right> \\
\left|P_4^{(0)}\right> \\
\left|P_5^{(0)}\right> \\
\left|P_6^{(0)}\right> \\
\left|P_{7+8}^{(0)}\right> \\
\left|P_9^{(0)}\right> \\
\left|P_{10}^{(0)}\right> \\
\left|P_{11}^{(0)}\right> \\
\left|P_{12}^{(0)}\right> \\
\left|P_{13}^{(0)}\right> \\
\left|P_{14}^{(0)}\right>
\end{array}
\end{equation}
with the basis states shown beside their corresponding rows in the notation of Sec.~\ref{sec:symmetries} and
the extra shorthand notation $\left|P_{7+8}^{(0)}\right>\equiv\tfrac{1}{\sqrt{2}}\left(\left|P_7^{(0)}\right>+\left|P_8^{(0)}\right>\right)$.
For $N_{\rm plaq}=2$ and $j_{\rm max}=1$, the vacuum block is
\begin{equation}\label{eq:N2j1}
H = \frac{g^2}{2}\left(
\begin{array}{cccccccccccccc}
0 & -2\sqrt{2}x & 0 & 0 & 0 & 0 & 0 & 0 & 0 & 0 & 0 & 0 & 0 & 0 \\
\!\!\!-2\sqrt{2}x\!\!\! & 3 & -\tfrac{x}{\sqrt{2}} & 0 & -\sqrt{3}x & 0 & -\tfrac{3x}{\sqrt{2}} & 0 & 0 & -2x & 0 & 0 & 0 & 0 \\
0 & -\tfrac{x}{\sqrt{2}} & 3 & 0 & 0 & -\sqrt{3}x & 0 & -\tfrac{3x}{\sqrt{2}} & 0 & 0 & 0 & 0 & 0 & 0 \\
0 & 0 & 0 & 4 & 0 & -\sqrt{\tfrac{8}{3}}x & 0 & 0 & 0 & 0 & 0 & 0 & 0 & 0 \\
0 & -\sqrt{3}x & 0 & 0 & 5 & \sqrt{2}x & 0 & -\tfrac{x}{\sqrt{3}} & 0 & 0 & 0 & 0 & 0 & 0 \\
0 & 0 & -\sqrt{3}x & -\sqrt{\tfrac{8}{3}}x & \sqrt{2}x & 5 & -\tfrac{x}{\sqrt{3}} & 0 & 0 & 0 & \!\!-\sqrt{\tfrac{8}{3}}x\!\! & 0 & -\tfrac{4}{3}x & 0 \\
0 & -\tfrac{3x}{\sqrt{2}} & 0 & 0 & 0 & -\tfrac{x}{\sqrt{3}} & 7 & -\tfrac{x}{9\sqrt{2}} & 0 & 0 & 0 & 0 & 0 & 0 \\
0 & 0 & -\tfrac{3x}{\sqrt{2}} & 0 & -\tfrac{x}{\sqrt{3}} & 0 & -\tfrac{x}{9\sqrt{2}} & 7 & -\tfrac{\sqrt{8}x}{3} & -\tfrac{2}{3}x & 0 & -\tfrac{8x}{3\sqrt{3}} & -\tfrac{4\sqrt{2}x}{3\sqrt{3}} & -\tfrac{8\sqrt{2}}{9}x\!\! \\
0 & 0 & 0 & 0 & 0 & 0 & 0 & -\tfrac{\sqrt{8}x}{3} & 8 & 0 & 0 & 0 & 0 & 0 \\
0 & -2x & 0 & 0 & 0 & 0 & 0 & -\tfrac{2}{3}x & 0 & 8 & 0 & 0 & 0 & 0 \\
0 & 0 & 0 & 0 & 0 & -\sqrt{\tfrac{8}{3}}x & 0 & 0 & 0 & 0 & 8 & 0 & 0 & 0 \\
0 & 0 & 0 & 0 & 0 & 0 & 0 & -\tfrac{8x}{3\sqrt{3}} & 0 & 0 & 0 & 10 & 0 & 0 \\
0 & 0 & 0 & 0 & 0 & -\tfrac{4}{3}x & 0 & -\tfrac{4\sqrt{2}x}{3\sqrt{3}} & 0 & 0 & 0 & 0 & 10 & 0 \\
0 & 0 & 0 & 0 & 0 & 0 & 0 & -\tfrac{8\sqrt{2}}{9}x & 0 & 0 & 0 & 0 & 0 & 12
\end{array}
\right)
\end{equation}
where the basis states, in order from top to bottom, are
\begin{equation}
\begin{array}{c}
\left|1_1^11_1^1\right> \,, \\
\tfrac{1}{\sqrt{2}}\left(\left|2_2^22_1^1\right>+\left|2_1^12_2^2\right>\right) \,, \\
\left|1_2^21_2^2\right> \,, \\
\tfrac{1}{\sqrt{2}}\left(\left|1_3^11_3^1\right>+\left|1_1^31_1^3\right>\right) \,, \\
\tfrac{1}{\sqrt{2}}\left(\left|3_2^21_2^2\right>+\left|1_2^23_2^2\right>\right) \,, \\
\tfrac{1}{2}\left(\left|2_3^12_2^2\right>+\left|2_2^22_3^1\right>+\left|2_1^32_2^2\right>+\left|2_2^22_1^3\right>\right) \,, \\
\left|3_2^23_2^2\right> \,, \\
\tfrac{1}{\sqrt{2}}\left(\left|2_2^22_3^3\right>+\left|2_3^32_2^2\right>\right) \,, \\
\left|1_3^31_3^3\right> \,, \\
\tfrac{1}{\sqrt{2}}\left(\left|3_3^33_1^1\right>+\left|3_1^13_3^3\right>\right) \,, \\
\tfrac{1}{\sqrt{2}}\left(\left|3_3^13_1^3\right>+\left|3_1^33_3^1\right>\right) \,, \\
\tfrac{1}{\sqrt{2}}\left(\left|1_3^33_3^3\right>+\left|3_3^31_3^3\right>\right) \,, \\
\tfrac{1}{2}\left(\left|3_1^33_3^3\right>+\left|3_3^33_1^3\right>+\left|3_3^13_3^3\right>+\left|3_3^33_3^1\right>\right) \,, \\
\left|3_3^33_3^3\right> \,.
\end{array}
\end{equation}
The presence of a positive off-diagonal entry in Eq.~(\ref{eq:N2j1}) suggests that the matrix is not perfectly suited for today's D-Wave hardware \cite{McGeoch}, but it does not represent any sign problem in the underlying theory and we can obtain accurate numerical results on D-Wave as discussed in Sec.~\ref{sec:eigenvalues}.

\end{widetext}

\section{THE AQAE ALGORITHM}\label{sec:AQAE}

The quantum annealer eigensolver was defined by Teplukhin {\it et al.} in Ref.~\cite{Teplukhin_2019} to solve the
eigenvalue problem for a real symmetric matrix on a quantum annealer.
It was later augmented with an automated search for $\lambda$ [the penalty parameter that appears for example in our Eq.~(\ref{eq:F})] \cite{TeplukhinNature} and was also extended to accommodate complex matrices \cite{TeplukhinPCCP}.
Classical algorithms for splitting larger QUBO problems into smaller ones for use with QAE were assessed in Ref.~\cite{teplukhin2021sampling}.
In the present work we introduce an adaptive search.

In our adaptive QAE algorithm, the standard QAE algorithm runs several times in sequence, with the best eigenvector from one run being used as input for
the next.
The original run searches within the set of vectors having real components $a_\alpha\in[-1,1)$ and it uses $K$ qubits to give $2^K$
equally spaced options for each $a_\alpha$.
Every subsequent run uses the same number of qubits but searches a range for each $a_\alpha$ that is half of the previous size,
so the allowed values are more finely spaced.
That smaller range is centered on the best value obtained from the previous run.
Our AQAE algorithm is outlined below, using some notation from the Supporting Information that accompanies Ref.~\cite{Teplukhin_2019}.

Begin with the vector set to zero, which is the center of the initial range $[-1,1)$,
\begin{equation}
a_\alpha^{(0)} = 0 \,.
\end{equation}
Now loop all remaining steps, i.e.\ Eqs.~(\ref{eq:beginQAE})-(\ref{eq:endQAE}), over the sequence of QAE runs while increasing the ``zoom factor'' $z$ for each subsequent run,
\begin{eqnarray}
{\rm for~} z = 0, 1, 2, \ldots && \\
C_\alpha^{(k)} &=& \left\{\begin{array}{cl} 2^{k-K-z}, & {\rm for~} k<K \\
                                            a_\alpha^{(z)}-2^{-z}, & {\rm for~} k=K \end{array}\right., \label{eq:beginQAE} \\
Q_{\alpha,n;\beta,m} &=& C_\alpha^{(n)}C_\beta^{(m)}(H_{\alpha\beta}-\lambda\delta_{\alpha\beta}). \\
F &=& \sum_{\alpha=1}^B\sum_{n=1}^K\sum_{\beta=1}^B\sum_{m=1}^K Q_{\alpha,n;\beta,m} q_n^\alpha q_m^\beta.\hspace{9mm} \\
&& {\rm Do~quantum~annealing.} \\
a_\alpha^{(z+1)} &=& \sum_{k=1}^KC_\alpha^{(k)}q_k^\alpha \label{eq:endQAE}
\end{eqnarray}
where $B$ is the number of rows in the matrix and $K$ is the number of logical qubits to be used.
The quantities $q_i$ are the binary variables that we introduced in Sec.~\ref{sec:eigenvalues}.
The quantity $F$ was defined in Eq.~(\ref{eq:F}) and is the dataset that gets provided directly to the D-Wave quantum annealer.
The iteration can be terminated after any quantum annealing step, for example when subsequent $z$ values are no longer providing an improved
(i.e.\ smaller) ground-state eigenvalue.

A \textsc{python} implementation of the AQAE algorithm is provided in Ref.~\cite{AQAE} with parameters set to reproduce Fig.~\ref{fig:adaptive}.

\begin{acknowledgments}
We are grateful to D-Wave Systems Inc.\ (Burnaby, Canada) for providing us with computing time on their quantum annealing hardware.
Our work was supported in part by a Discovery Grant from the Natural Sciences and Engineering Research Council (NSERC) of Canada.
S.A.\ received additional funding from a Dean's Undergraduate Research Award,
E.M.\ from a Carswell Graduate Scholarship, and S.P.\ from an NSERC Undergraduate Student Research Award.
\end{acknowledgments}

\bibliography{SU2bib}

\end{document}